\documentclass[]{interact}

\usepackage{epstopdf}
\usepackage[caption=false]{subfig}

\usepackage[numbers,sort&compress]{natbib}
\bibpunct[, ]{[}{]}{,}{n}{,}{,}
\renewcommand\bibfont{\fontsize{10}{12}\selectfont}

\theoremstyle{plain}

\theoremstyle{definition}

\theoremstyle{remark}

\usepackage[numbers]{natbib}

\usepackage{amsmath,amssymb,latexsym, amsfonts,amsthm}
\usepackage{float}
\usepackage[font=footnotesize]{caption}
\usepackage{color}
\usepackage{multirow}

\begin{document}

\articletype{ARTICLE TEMPLATE}

\title{Length-of-stay times in hospital for COVID-19 patients using the smoothed Beran's estimator with bootstrap bandwidth selection}

\author{
\name{Rebeca Peláez\textsuperscript{a}\thanks{CONTACT Rebeca Peláez. Email: rebeca.pelaez@uc3m.es}, Ricardo Cao, Juan M. Vilar\textsuperscript{b}}
\affil{\textsuperscript{a}Department of Statistics,
	Universidad Carlos III de Madrid, Madrid, Spain; \textsuperscript{b}Department of Mathematics,
	Universidade da Coruña, A Coruña, Spain}
}

\maketitle

\begin{abstract}
The survival function of length-of-stay in hospital ward and ICU for COVID-19 patients is studied in this paper. Flexible statistical methods are used to estimate this survival function given relevant covariates such as age, sex, obesity and chronic obstructive pulmonary disease (COPD). A doubly-smoothed Beran's estimator has been considered to this aim. The bootstrap method has been used to produce new smoothing parameter selectors and to construct confidence regions for the conditional survival function. Some simulation studies show the good performance of the proposed methods.
\end{abstract}


\section{Introduction}
\label{Introduction}

In biomedical studies, predicting a patient's survival time, $T$, given a covariate, $X$, related to the time (sex, cholesterol, age...) is a problem of interest. 
For patients that are in hospital, some event of interest is the time when they leave hospital ward or when they leave Intensive Care Unit (ICU), if they were admited to it. In those cases, the survival times are length-of-stay times in hospital, $T$, ward or ICU. 
In this scenario, the aim is to estimate the distribution function of $T$ conditional to $X=x$, that is, $F(t \vert x)=P(T \leq t \vert X=x)$ or, equivalently, to estimate the conditional survival function $S(t \vert x)= 1- F(t \vert x)$.
In many such situations, the variable $T$ is subject to right random censoring. Right censoring occurs when a proportion of survival times are unknown because the study ends before all individuals have experienced the event of interest.

The most commonly used nonparametric estimator of $F(t \vert x)$ under censoring was introduced by \cite{Beran1981}. This estimator  turns out to be the Kaplan-Meier estimator (see \cite{KaplanMeier1958}) in absence of covariates. Asymptotic properties of this estimator have been widely studied in the literature (\cite{Dabrowska1989, Manteiga1994, Ingrid1996, Iglesias1999}). Another nonparametric estimator of the conditional distribution function with censored data was proposed by \cite{Ingrid1999} and \cite{Ingrid2001}. This estimator presents a better behaviour than Beran's estimator when estimating the distribution function in the right tail with heavy censoring. In \cite{Gannoun2005} and \cite{Gannoun2007} an alternative estimator based on the local linear method proposed in \cite{Cai2003} is studied. All these nonparametric distribution estimators are based on a covariate smoothing.  
In a recent paper by \cite{Pelaez2022_JNPS} a general estimator of $S(t|x)$ based on a double smoothing both in the covariate and in the time variable is proposed. The idea presented there can be applied to any nonparametric estimator of the conditional survival function.
In \cite{Pelaez2022_JNPS}, simulation studies show how well this estimator performs when the objective is to estimate $S(t|x)$ for a fixed value of the covariate, $x$, and $t$ covering the interval $I_T \subset \mathbb{R}^+$. 
Asymptotic expressions for the bias and variance of the estimators were also proved in \cite{Pelaez2022_JNPS}, but there is no available method to choose the smoothing parameters involved.

Previous work has shown that bootstrap techniques shed light in this context. 
In \cite{Efron1981}, bootstrap for right censored data is firstly proposed and asymptotic theory was established by \cite{Akritas1986} and \cite{Lo_Singh1986}. Bootstrap for nonparametric regression with right censored observations at fixed covariate values is studied in \cite{Ingrid1997}. A bootstrap approach for the nonparametric censored regression setup is studied in \cite{Li_Datta2001}. In \cite{Geerdens2017} a local cross-validation bandwidth selector is proposed.

The goal of this work is to define a resampling technique to approximate the smoothing parameters involved in Beran's survival estimator and the smoothed Beran's survival estimator. 
Our proposal is based on combining the smoothed bootstrap with a weighted bootstrap for covariates. This approach follows the ideas of \cite{Li_Datta2001} for the conditional distribution function and \cite{Pelaez2022_Mathematics} for the probability of default function.
Pointwise theoretical confidence intervals derived from asymptotic theory for Beran's survival estimator and the smoothed Beran's survival estimator (see \cite{Iglesias1999} and \cite{Pelaez2022_JNPS}, respectively) are not computable in practice. Therefore, the bootstrap is also useful to compute confidence intervals and regions.

The remainder of this paper is organized as follows. In Section \ref{Bandwidth_selection}, bootstrap selectors for the bandwidths of Beran's and the smoothed Beran's estimators are proposed. In Section \ref{Bandwidth_Simulation} a simulation study shows the behaviour of the survival estimators with bootstrap bandwidths.
The issue of obtaining confidence regions for the conditional survival function, $S(t|x)$, for a fixed value of $x\in I \subseteq \mathbb{R}$ and $t$ covering some interval $I_T \subseteq \mathbb{R}^+$, is addressed using Beran's and the smoothed Beran's estimators in Section \ref{Confidence_regions}.
A simulation study on the proposed bootstrap methods for the calculation of confidence regions is shown in Section \ref{Confidence_Simulation}.
These methods are used in Section \ref{Real_data} to construct nonparametric estimations of length-of-stay in hospital ward and ICU for COVID-19 patients in Galicia, Spain, during the first weeks of the breakdown.

\section{Bandwidth selection for Beran's and the smoothed Beran's estimators}
\label{Bandwidth_selection}

Let $\{ (X_i, Z_i, \delta_i)\}_{i = 1}^n$ be a simple random sample of $(X, Z, \delta)$ with $X$ being the covariate, $Z = \min \{ T, C \}$ the observed variable and $\delta = I(T\leq C)$ the uncensoring indicator, with $T \geq 0$ and $C \geq 0$ represent the life and censoring times.
The distribution function of $T$ is denoted by $F(t)$ and the survival function by $S(t) = 1 - F(t)$. The functions $F(t|x)=P(T \leq t|x)$ and 
$S(t|x)=P(T>t|x)$ are the distribution and survival functions of $T$ evaluated at $t$ conditional to $X  = x$. The conditional distribution function of $Z$ is denoted by $H(t|x)$ and the conditional distribution function of $C$ is denoted by $G(t|x)$. The random variables $T$ and $C$ are conditionally independent given $X = x$.

In this section, methods for the automatic selection of the bandwidths for Beran's estimator and the smoothed Beran's estimator of the conditional survival function are proposed. 

\subsection{Beran's estimator}
\label{Bandwidth_Beran}

In \cite{Beran1981}, the generalisation of the product-limit estimator for the conditional survival function is proposed and given by
\begin{equation}
	\label{eq:S_Beran}
	\begin{array}{ccl}
		\widehat{S}_h(t \vert x) & = & \displaystyle\prod_{i = 1}^{n} \bigg( 1 - \dfrac{I(Z_i \leq t, \; \delta_i = 1) w_{h, i}(x)}{ 1 - \sum_{j=1}^{n} I(Z_j < Z_i) w_{h, j}(x) }\bigg)
	\end{array}
\end{equation}
where $ w_{h, i}(x) = \dfrac{K\big((x - X_i)/ h\big)}{ \sum_{j = 1}^{n} K\big((x - X_j) / h\big) } $
with $i = 1, ..., n$,  $K(u)$ is a kernel function and $h=h_n$ is the bandwidth
that indicates the degree of smoothing introduced into the estimator through the covariate.

In \cite{Pelaez2022_Mathematics}, an algorithm for bootstrap resampling based on Beran's estimator is proposed to obtain resamples ${(X_i^*, Z_i^*, \delta_i^*)}_{i=1}^n$  of the sample ${(X_i, Z_i, \delta_i)}_{i=1}^n$. This resampling method is used to approximate the smoothing bandwidth required to estimate the probability of default, a certain curve of interest in credit risk contexts (see the works of \cite{Pelaez2021_TEST}, \cite{Pelaez2021_SORT}). The detailed algorithm for bootstrap resampling based on Beran's estimator and explanations on its construction can be found in \cite{Pelaez2022_Mathematics}. 
In this paper, we consider this resampling algorithm to approximate the bootstrap bandwidth for Beran's estimator of the conditional survival curve.

In order to estimate the conditional survival function, $S(t|x)$, for a fixed $x\in I$ and $t$ covering the interval $I_T \subset \mathbb{R}$, our benchmark is the bandwidth $h_{MISE} \in I_1$, that minimizes the mean integrated squared error given by
\begin{equation}
	\label{eq:S_MISE_Beran}
	MISE_x(h) = E\bigg( \int_{I_T} \big( \widehat{S}_h(t|x) - S(t|x) \big)^2 dt  \bigg).
\end{equation}
Consider the bootstrap resample ${(X_i^*, Z_i^*, \delta_i^*)}_{i=1}^n$ generated by the resampling techniqued from \cite{Pelaez2022_Mathematics}.
The bootstrap approximation of the function $MISE_x(h)$ is given by
$$ MISE_x^*(h) = E^*\bigg( \int_{I_T} \big( \widehat{S}_h^{*}(t|x) - \widehat{S}_r(t|x) \big)^2 dt  \bigg)$$
where $\widehat{S}_r(t|x)$  is the estimation of the theoretical survival function with pilot bandwidth, $r$, using the sample $\big\{(X_i, Z_i, \delta_i)\big\}_{i=1}^{n}$ and $ \widehat{S}_h^{*}(t|x)$ is the bootstrap estimation of $S(t|x)$ with bandwidth $h$, using the bootstrap resample $\big\{(X_i^*, Z_i^*, \delta_i^*)\big\}_{i=1}^{n}$.

The resampling distribution of $\widehat{S}_h^*(t|x)$ cannot be computed in a closed form, so the Monte Carlo method is used. It is based on obtaining $B$ bootstrap resamples and estimating $\widehat{S}_h^{*}(t|x)$ for each of them. Thus, the distribution of $\widehat{S}_h^{*}(t|x)$ is approximated by the empirical one of $\widehat{S}_h^{*, 1}(t|x), \dots, \widehat{S}_h^{*, B}(t|x)$, obtained from $B$ bootstrap resamples and the Monte Carlo approximation of the bootstrap MISE is given by
\begin{equation}
	\label{eq:S_MISE_Beran_boot}
	{\emph{MISE}}_x^*(h) \simeq  \dfrac{1}{B} \sum_{k=1}^B  \bigg( \int_{I_T} \big( \widehat{S}_h^{*, k}(t|x) - \widehat{S}_r(t|x) \big)^2 dt\bigg).
\end{equation} 
Likewise, the integral is approximated by a Riemann sum.

\subsubsection*{Algorithm for bootstrap bandwidth selection for Beran's estimator}

Let $x\in I$ be a fixed value of the covariate and $r \in I_1$:

\begin{enumerate}
	\item Compute $\widehat{S}_{r}(t |x)$ from the original sample $\{(X_i, Z_i, \delta_i)\}_{i=1}^n$ for some values of $t$ in a grid of $I_T$.
	
	\item Obtain $B$ bootstrap resamples of the form $\{(X_i^{*, k}, Z_i^{*, k}, \delta_i^{*, k})\}_{i=1}^n$ with $k = 1, ..., B$ using the bootstrap based on Beran's estimator with pilot bandwidth $r\in I_1$ and calculate $\widehat{S}^{*, k}_h(t |x)$ for each of them, for the same values of $t$ in the grid of $I_T$.
	
	\item {Approximate} $MISE_x^*(h)$ according to \eqref{eq:S_MISE_Beran_boot}.
	
	\item Repeat Steps 1--3 for values of $h$ in a grid of $I_1$.
	
	\item Select the value of $h$ that provides the smallest $MISE_x^*(h)$ as the bootstrap bandwidth $h^*$.
\end{enumerate}

Concerning the auxiliary bandwidth, $r\in I_1$, preliminary analyses suggests the following choice as a suitable pilot bandwidth in this context:
\begin{equation}
	\label{eq:Bootstrap_S_piloto_r}
	r = c\dfrac{\big(  Q_{X}(0.975) - Q_{X}(0.025) \big)}{2}  \bigg(   \sum_{i=1}^n \delta_i\bigg)^{-1/3}
\end{equation}
where $Q_X (u)$ is the $u$ quantile of the sample $\big\{X_i\big\}_{i=1}^n$.
Equation \eqref{eq:Bootstrap_S_piloto_r} considers the variability of the covariate, $ Q_{X}(0.975) - Q_{X}(0.025)$, and the uncensored sample size, $\sum_{i=1}^n \delta_i$. The exponent of this sample size, $-1/3$, is typically appropriate in selection of the optimal bandwidth for estimating the distribution function (see \cite{Azzalini1981} and \cite{Jones1990}). This expression was derived after several attempts in the simulation studies. These analyses show that choosing $c<1$ increases the estimation error of Beran's estimator since the bootstrap method provides excessively small bandwidths. In general, $c \geq 1$ is considered, with the choice $c = 3/2$ being appropriate. In cases where the function $E(T|X = x)$ is found to be highly variable with respect to $x$, smaller bandwidths may be considered and our proposal there is $c = 1$.

Note that the proposed algorithm is also valid to obtain a bootstrap approximation of the optimal bandwidth for the estimation of $S(t|x)$ for fixed values of $t\in I_T$ and $x \in I$ by replacing $MISE_x^*(h)$ by $MSE_{t, x}^*(h)$, which is the bootstrap analogue of 
$$ MSE_{t,x}(h) = E\bigg( \big( \widehat{S}_h(t|x) - S(t|x) \big)^2  \bigg).$$

\subsection{The smoothed Beran's estimator}
\label{Bandwidth_SBeran}

Nonparametric estimators of the conditional survival function such as Beran's estimator defined in \eqref{eq:S_Beran} are smoothed just on the covariate $X$. However, time variable smoothing of the conditional survival estimators  has been found to be useful for the graphical representation, as well as to reduce the estimation error. This smoothing in both the covariate and the time variable was previously used in \cite{Foldes1981} and \cite{Leconte2002}.
Recently, the smoothed Beran's estimator of the conditional survival function was proposed and studied in \cite{Pelaez2022_JNPS}. Simulation studies have shown that this doubly smoothed estimator performs better than classical survival estimators smoothed only in the covariate. The smoothed Beran's survival estimator is defined by
\begin{equation}
	\begin{array}{ccl}
		\widetilde{S}_{h, g} (t \vert x) & = & 1 - \displaystyle \sum_{i=1}^{n} s_{(i)} \mathbb{K} \bigg(\dfrac{t - Z_{(i)}}{g}\bigg)
		\label{eq:S_smoothed_estimator}
	\end{array}
\end{equation}
where $s_{(i)} = \widehat{S}_h(Z_{(i-1)}\vert x) -  \widehat{S}_h(Z_{(i)}\vert x)$ with $Z_{(i)}$ the $i$-th element of the sorted sample of $Z$, $\widehat{S}_h(t\vert x)$ is Beran's estimator given in \eqref{eq:S_Beran}, $\mathbb{K}(t)$ is the distribution function of a kernel $K$, $\mathbb{K}(t) = \int_{- \infty}^{t} K(u)du$, and $g = g_{n}$ is the smoothing parameter for the time variable.

Asymptotic expressions for the bias and variance of this estimator are just too complex to calculate plug-in bandwidths (see \cite{Pelaez2022_JNPS}). Moreover, no method is available for choosing the smoothing parameters involved in this estimator. Therefore, this paper proposes a bootstrap selector for the two-dimensional bandwidths $(h, g)$.

In \cite{Pelaez2022_Mathematics} a weighted bootstrap with covariates combined with a smoothed bootstrap is proposed to obtain resamples ${(X_i^*, Z_i^*, \delta_i^*)}_{i=1}^n$  of the sample ${(X_i, Z_i, \delta_i)}_{i=1}^n$. This resampling method is used to approximate the smoothing bandwidths needed to estimate the probability of default curve. In this paper, we consider this resampling algorithm to approximate the bootstrap bandwidths for the smoothed Beran's estimator of the conditional survival function. The detailed algorithm for bootstrap resampling based on the smoothed Beran's estimator and explanations on its construction can be found in \cite{Pelaez2022_Mathematics}.

Under this resampling plan, the bootstrap conditional distribution function of $T^*|X^*$ is the smoothed Beran's estimator,  $\widetilde{F}_{r, s}(t|X_i^*)$. Similarly, the bootstrap conditional distribution function of $C^*|X^*$ is $\widetilde{G}_{r, s}(t|X_i^*)$.

The optimal bivariate bandwidth, $(h_{MISE}, g_{MISE})\in I_1\times I_2$ is defined as the pair of bandwidths that minimizes the mean integrated squared error given by
\begin{equation}
	\label{eq:S_MISE_SBeran}
	MISE_x(h,g) = E \left( \displaystyle \int_{I_T} \big( \widetilde{S}_{h,g}(t\vert x) - S(t\vert x) \big)^2 dt \right).
\end{equation}
The bootstrap version of $MISE_x(h,g)$ is given by 
$$ MISE_x^*(h, g)  =  E^*  \bigg( \int_{I_T} \big(\widetilde{S}_{h, g}^{*}(t|x) - \widetilde{S}_{r, s}(t|x) \big)^2 dt\bigg),$$
where $\widetilde{S}_{r, s}(t|x) $ is the smoothed Beran's survival estimation with pilot bandwidths $(r, s) \in I_1 \times I_2$ using the sample $\big\{(X_i, Z_i, \delta_i)\big\}_{i=1}^{n}$ 
and $\widetilde{S}_{h, g}^{*}(t|x)$ is the bootstrap estimation of $S(t|x)$ with bandwidths $(h, g)$, using the bootstrap resample $\big\{(X_i^*, Z_i^*, \delta_i^*)\big\}_{i=1}^{n}$.
Since the resampling distribution of $\widetilde{S}_{h, g}^{*}(t|x)$ cannot be computed in a closed form, the Monte Carlo method is used. 
The bootstrap bandwidth selection algorithm is defined as follows.

\subsubsection*{Algorithm for bootstrap bandwidth selection for the smoothed Beran's estimator}

Let $x$ be a fixed value of the covariate and $(r, s) \in I_1\times I_2$: 

\begin{enumerate}
	\item Compute $\widetilde{S}_{r, s}(t |x)$ from the original sample $\{(X_i, Z_i, \delta_i)\}_{i=1}^n$, for some values of $t$ in a grid of $I_T$.
	
	\item Obtain $B$ bootstrap resamples of the form $\{(X_i^{*, k}, Z_i^{*, k}, \delta_i^{*, k})\}_{i=1}^n$ with $k = 1, ..., B$ using the bootstrap based on the smoothed Beran's estimator and calculate $\widetilde{S}^{*, k}_{h, g}(t |x)$ for each of them, for the same values of $t$ in the grid of $I_T$.
	
	\item Approximate $\emph{MISE}_x^*(h, g)$ according to
	\begin{equation}
		\label{eq:S_MISE_SBeran_boot}
		{\emph{MISE}}_x^*(h, g) \simeq  \dfrac{1}{B} \sum_{k=1}^B  \bigg( \int_{I_T} \big( \widetilde{S}_{h, g}^{*, k}(t|x) - \widetilde{S}_{r, s}(t|x) \big)^2 dt\bigg).
	\end{equation}
	
	\item Repeat Steps 1--3 for pairs of values $(h, g)$ in a grid of $I_1\times I_2$.
	
	\item Obtain the pair $(h, g)$ that provides the smallest $MISE_x^*(h, g)$ as the bootstrap bandwidth $(h^*, g^*)$.
\end{enumerate}

The auxiliary bandwidth $r \in I_1$ is chosen as in \eqref{eq:Bootstrap_S_piloto_r}. The pilot bandwidth $s \in I_2$ for the time variable smoothing is chosen using the following formula
\begin{equation}
	\label{eq:Bootstrap_S_piloto_s}
	s = \dfrac{3}{4} \big(  Q_{Z}(0.975) - Q_{Z}(0.025)   \big) \bigg(   \sum_{i=1}^n \delta_i\bigg)^{-1/7}
\end{equation}
where $Q_Z (u)$ is the $u$ quantile of the sample $\big\{Z_i\big\}_{i=1}^n$.
This expression was derived after several attempts in the simulation studies. It takes into account the variability of the observed time variable, $ Q_{Z}(0.975) - Q_{Z}(0.025)$, and the sample size of the uncensored population, $\sum_{i=1}^n \delta_i$. The exponent of this sample size, $-1/7$, is heuristically deduced from the asymptotic expression of the MISE of the survival estimators proved in \cite{Pelaez2022_JNPS}.

\section{Simulation study for bandwidth selection} 
\label{Bandwidth_Simulation}

A simulation study is conducted in order to show the behaviour of the bootstrap bandwidth selectors for Beran's and smoothed Beran's estimators proposed in Section \ref{Bandwidth_selection}. 
Two models are considered, one based on Weibull distributions for life and censoring times and the other on exponential distributions.  The data generation models were previously used in \cite{Pelaez2022_Mathematics} in order to analyse the associated default probability curves. In this paper they are used to discuss the bandwidth selection algorithms presented above.

Model 1 considers a $U(0, 1)$ distribution for $X$. The time to occurrence of the event conditional to the covariate, $T\vert_{X = x}$,  follows a Weibull distribution with parameters 2 and $A(x)^{-1/2}$ and $A(x)= 1 + 5x$. The censoring time conditional to the covariate, $C\vert_{X = x}$, follows a Weibull distribution with parameters $2$ and $B(x)^{-1/2}$ and $B(x) = 10 + d_1x + 20x^2$. In this case, the conditional survival function and the censoring conditional probability are given by:
\begin{align*}
S(t \vert x) & =  e^{-A(x)t^2}, \\
P(\delta= 0 \vert X=x ) & =  \dfrac{B(x)}{A(x) + B(x)}.
\end{align*}
Having set the covariate value, $x = 0.6$, the values of $d_1$ are chosen so that the censoring conditional probability is $0.2$ and $0.5$. These values are $d_1 = -27$ and $d_1 = -22$, respectively. The conditional survival function for this model is estimated in a time grid of size $n_T$, $0 < t_1 < \cdots< t_{n_T}$, where $t_{n_T} = F^{-1}(0.95|x) = 0.8654$ for the covariate value $x = 0.6$.  Therefore, in this case $I_T=(0, 0.8654)$.

Model 2 considers a $U(0, 1)$ distribution for $X$. The time to occurrence of the event conditional to the covariate, $T\vert_{X = x}$, follows an exponential distribution with parameter $\Gamma(x)= 2 + 58x -160x^2 + 107x^3$. The censoring time conditional to the covariate, $C\vert_{X = x}$, follows an exponential distribution with parameter 
$\Delta(x)= 10 + c_1x + 20x^2$. In this scenario, the conditional survival function and the censoring conditional probability are the following:
\begin{align*}
S(t \vert x) & = e^{-\Gamma(x)t}, \\
P(\delta= 0 \vert X=x ) & = \dfrac{\Delta(x)}{\Gamma(x) + \Delta(x)}. 
\end{align*}
Having set the covariate value, $x = 0.8$, the values of $c_1$ are chosen so that the censoring conditional probability is $0.2$ and $0.5$. These values are $c_1=-113/4$ and $c_1=-55/2$, respectively. The conditional survival function is estimated in a time grid of size $n_T$, $0 < t_1 < \cdots < t_{n_T}$, where $t_{n_T} = F^{-1}(0.95|x) = 3.8211$ for the covariate value $x = 0.8$. Therefore, in this case $I_T=(0, 3.8211)$.

It can be proved that Model 1 is close to a proportional hazards model, while Model 2 moves away from this parametric model with the aim of widely considering different simulation scenarios. Although in both models the distribution of the covariate is chosen to be uniform, it was previously verified that the distribution of $X$ has no effect on the behaviour of the analysed methods. The boundary effect is corrected using the reflexion principle (see \cite{Billingsley1968}) and the truncated Gaussian kernel with a truncation range $(-50, 50)$ is used. The size of the lifetime grid is $n_T = 100$.  The sample size is $n = 400$.
Regarding the auxiliary bandwidths, Model 1 uses $c = 3/2$, while Model 2 considers $c = 1$. The reason for this choice is that the distribution of the time variable in Model 2 varies much more with respect to $x$, thus requiring smaller bandwidths.

\subsection{Simulation study for Beran's estimator}
\label{Bandwidth_Simulation_Beran}

In this subsection, the behaviour of the bootstrap bandwidth selector for Beran's estimator is analysed.
For each model, the estimation error function $MISE_{x}(h)$ is approximated via Monte Carlo using $300$ simulated samples. The bandwidth that minimises $MISE_{x}(h)$ is obtained and denoted by $h_{MISE}$. The values of $h_{MISE}$ and  $MISE_{x}(h_{MISE})$ are used as a benchmark.

In the simulation study, $N = 300$ simulated samples are used to evaluate the performance of the bootstrap bandwidth selector. For each sample, $B = 500$ bootstrap resamples are generated to approximate the bootstrap MISE function, $MISE_x^*(h)$, and obtain the bootstrap bandwidth associated to each simulated sample, $h_{j}^{\ast }$, $j=1,2,\ldots, N$. The mean value of the $N$ bootstrap bandwidths and their standard deviation are defined as
follows 
$$  \overline{h^*} = \dfrac{1}{N} \sum_{j=1}^N  h^*_{j}, \quad   sd\big(  h^* \big) = \sqrt{ \dfrac{1}{N}  \sum_{j=1}^{N}   \big( h^*_{j}   - \overline{h^*}  \big)^2  }.  $$

In order to minimise the error function $MISE$ without increasing CPU time more than necessary, a limited-memory algorithm for solving large nonlinear optimization problems is used, L-BFGS-B. It was proposed by \cite{optim1995} for solving optimization problems subject to simple bounds on the variables in which information on the Hessian matrix is difficult to obtain. Results of numerical studies about this method are shown in \cite{optim1995}. It is available at the \textsl{stats} package from the Comprehensive R Archive Network (CRAN). It uses Fortran 77 subroutines (see \cite{optim1997}).

A relative measure of the difference between the bootstrap bandwidth and the optimal one, is considered
$$  H^*_{j}= \dfrac{h^*_{j} - h_{MISE}}{h_{MISE}}, $$
with $j = 1, \dots, N$. The mean of the absolute value of these relative deviations, $\overline{H^*} =  \frac{1}{N} \sum_{j = 1}^{N} |H^*_{j}| $, is a good measure of how close the bootstrap bandwidth is to the optimal one.

For each sample, the estimation error of Beran's estimator with the corresponding bootstrap bandwidth,
$$ \displaystyle MISE_x(h^*_{j}) =  E\bigg(  \int_{I_T} \big( \widehat{S}_{h^*_{j}}(t|x) -S(t|x) \big)^2 dt  \bigg), $$
and its square root, $RMISE_x(h^*_{j})$, are approximated via Monte Carlo using $300$ simulated samples. The mean of these estimation errors given by 
$$ \overline{RMISE_x(h^*) }  = \dfrac{1}{N} \sum_{j = 1}^N RMISE_x(h^*_{j})$$
is used as a measure of the estimation error made by the bootstrap bandwidth, when compared with the estimation error made by the MISE bandwidth.

As a relative measure of the difference between the estimation errors using the bootstrap and the MISE bandwidths, the following ratios are defined:
$$  R^*_{j} =  \dfrac{RMISE_x(h^*_{j}) - RMISE_x(h_{MISE})}{RMISE_x(h_{MISE})}$$
satisfying $ R^*_{j} \geq 0$  for all $j= 1,\dots, N$. The mean of the $R_{j}^{\ast }$ values with $j = 1, \dots, N$ is denoted by $ \overline{R^*} = \frac{1}{N}\sum_{j = 1}^{N} R^*_{j}$.
Small values (close to zero) of $\overline{H^{\ast }}$ and $\overline{R^{\ast }}$ indicate good behavior of the bootstrap bandwidth.

\begin{figure}[H]
	\centering
	\includegraphics[width=0.4\linewidth]{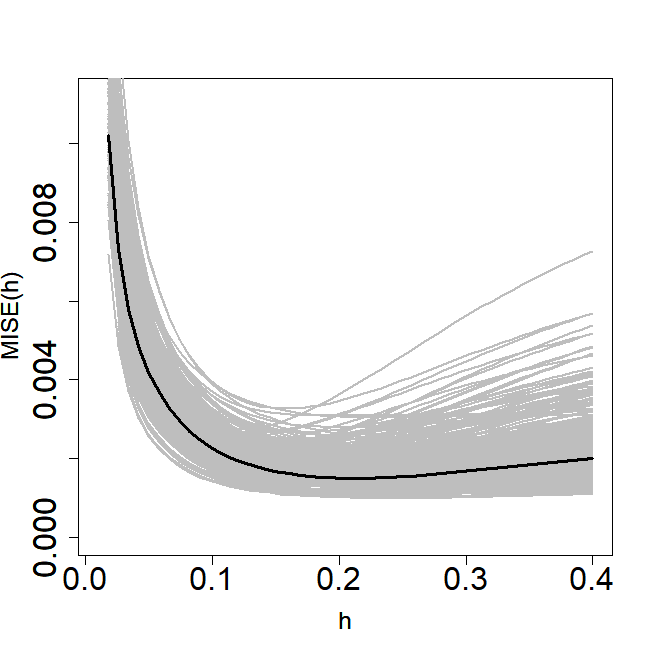}
	\includegraphics[width=0.4\linewidth]{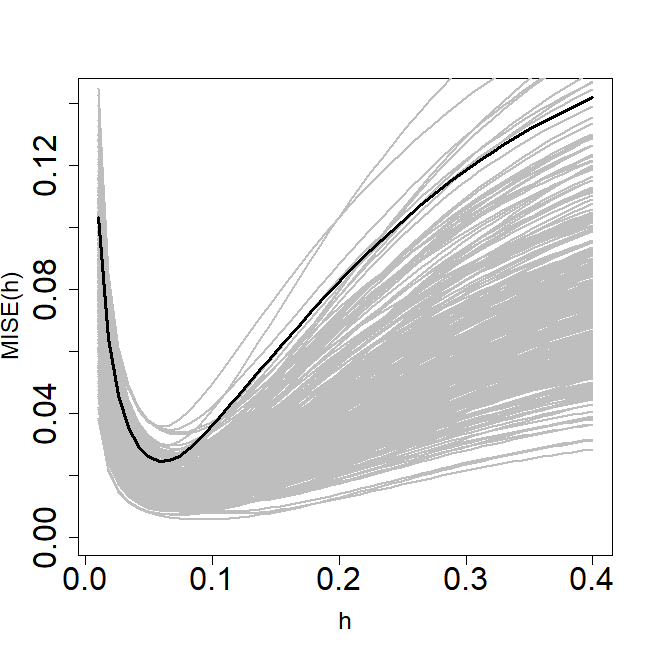}
	
	\caption{$MISE_x(h)$ function (black line) approximated via Monte Carlo and $MISE_x^*(h)$ functions (gray lines) for $N=300$ samples when $P(\delta = 0|x) = 0.5$ in Model 1 (left) and Model 2 (right).}
	\label{fig:Bootstrap_S_MISE_MISEboot_Beran}
\end{figure}

Figure \ref{fig:Bootstrap_S_MISE_MISEboot_Beran} shows the MISE function and the bootstrap MISE approximations for Models 1 and 2. Values of the bootstrap bandwidths, estimation errors and relative measures for Models 1 and 2 are included in Table \ref{tab:Bootstrap_S_Bandwidths_RMISE_Beran}. 
The results show a good performance of the proposed bootstrap selector, especially, in Model 1. 
Table \ref{tab:Bootstrap_S_Bandwidths_RMISE_Beran} shows that using bootstrap bandwidth selection increases just around $4\%$ to $6\%$ the RMISE error with respect to the optimal bandwidths for Model 1. For Model 2, the increase in the RMISE error caused by bootstrap bandwidth selection is around $11\%$ to $15\%$.

\begin{table}[H]
	\centering
	\renewcommand{\arraystretch}{1}
		\begin{tabular}{c|c|c|c|c|}
			\cline{2-5}
			& \multicolumn{2}{c|}{Model 1} &   \multicolumn{2}{c|}{Model 2}  \\ \hline
			\multicolumn{1}{|c|}{$P(\delta = 0|X=x)$} & $0.2$ & $ 0.5$  & $ 0.2$ & $ 0.5$ \\ \hline
			\multicolumn{1}{|c|}{ $h_{MISE}$ }
			& 0.23939 & 0.21212 & 0.04515 & 0.05687 \\
			\multicolumn{1}{|c|}{ ${RMISE_x(h_{MISE})}$ }
			& 0.02411 & 0.03652 & 0.11612 & 0.15576  \\ \hline
			\multicolumn{1}{|c|}{ $\overline{h^*} \quad (sd)$ }
			& 0.23815 (0.093) & 0.21897 (0.082) & 0.06718 (0.007)  & 0.08082 (0.011)  \\	
			\multicolumn{1}{|c|}{ ${\overline{H^*}}$}
			& 0.29033 & 0.26199  & 0.48794 & 0.42119 \\
			\multicolumn{1}{|c|}{${\overline{RMISE_x(h^*)}}$ } 
			& 0.02548 & 0.03809 & 0.13391 & 0.17242 \\	
			\multicolumn{1}{|c|}{${\overline{R^*}}$ }   
			& 0.05762 & 0.04373 & 0.15316 & 0.10698 \\ \hline
			
		\end{tabular}
	\caption{MISE and average bootstrap bandwidths and estimation errors of Beran's survival estimator in each level of conditional censoring probability for Models 1 and 2. Numbers within brackets are standar deviations.}
	\label{tab:Bootstrap_S_Bandwidths_RMISE_Beran}
\end{table}

\subsection{Simulation study for the smoothed Beran's estimator}
\label{Simulation_SBeran}

In this section, a simulation study on the bootstrap bandwidth selector for the smoothed Beran's estimator in \eqref{eq:S_smoothed_estimator} is carried out. The resampling technique and Monte Carlo approximation of the MISE presented in Section \ref{Bandwidth_SBeran} are used. Models 1 and 2 are considered.

For each model, the error function $MISE_{x}(h, g)$ is approximated via Monte Carlo from 300 simulated samples and the bivariate bandwidth that minimises $MISE_{x}(h, g)$ is obtained and denoted by $(h_{MISE}, g_{MISE})$. The values of $(h_{MISE}, g_{MISE})$ and $MISE_{x}(h_{MISE}, g_{MISE})$ are used as a benchmark.

In the study, $ N = 300$ samples are simulated to evaluate the performance of the bootstrap bandwidth selector. For each simulated sample, the corresponding bootstrap bandwidths are approximated from $B=500$ resamples, obtaining $(h_j^*,  g_j^*)$ with $j = 1, \dots, N$. The mean value of the $N$ bootstrap bandwidths and their standard deviation are the following:
$$ (\overline{h^*}, \overline{g^*}) = \bigg(     \dfrac{1}{N} \sum_{j=1}^N  h^*_{j} ,  \dfrac{1}{N} \sum_{j=1}^N  g^*_{j}     \bigg), $$
$$ sd\big(  h^* \big) = \sqrt{\dfrac{1}{N} \sum_{j=1}^{N} \big( h^*_{j}   - \overline{h^*} \big)^2   }, \quad  sd\big(  g^* \big) =  \sqrt{\dfrac{1}{N} \sum_{j=1}^{N} \big( g^*_{j}   - \overline{g^*} \big) ^2  } . $$
A limited-memory algorithm for solving large nonlinear optimization problems, L-BFGS-B, is used to minimise the error function $MISE$, as explained in Section \ref{Bandwidth_Simulation_Beran}.

In order to measure the distance of the bootstrap two-dimensional bandwidth of the $j$-th sample, $(h_j^* , g_j^*)$, from the corresponding MISE bandwidth, $(h_{MISE}, g_{MISE})$, the two-dimensional vector was consider:
$$ D^*_j = \left(\dfrac{h_j^* - h_{MISE}}{h_{MISE}},  \dfrac{g_j^* - g_{MISE}}{g_{MISE}} \right)$$
and its Euclidean norm denoted by $ H_j^* = \| D^*_j   \|_2$ with $j = 1, \dots, N$. The mean value, 
$\overline{H^*} = \frac{1}{N} \sum_{j = 1}^{N} H_j^*$
is a measure of how close the bootstrap bandwidths are to the MISE one.

For each sample, the estimation error of the smoothed Beran's estimator with the corresponding bootstrap bandwidth,
$$ \displaystyle MISE_x(h_j^*, g_j^*) =  E\bigg(  \int_{I_T} \big( \widetilde{S}_{h_j^*, g_j^*}(t|x) - S(t|x) \big)^2 dt  \bigg), $$
and its square root, $RMISE_x(h_j^*, g_j^*)$, are approximated via Monte Carlo using $300$ simulated samples.
The mean of these estimation errors given by
$$ \overline{RMISE_x(h^*, g^*)} = \dfrac{1}{N} \sum_{j=1}^N RMISE_x\big(h^*_{j}, g^*_{j}\big)$$
is used as a measure of the estimation error made by the bootstrap two-dimensional bandwidth in the model.

The ratio
$$ {R_j^*} = \dfrac{  {RMISE_x(h_j^*, g_j^*)}  - RMISE_x(h_{MISE}, g_{MISE})}{RMISE_x(h_{MISE}, g_{MISE})}$$
is defined as a relative measure of the difference between the error of the estimator with bootstrap bandwidth and MISE bandwidth. 
The mean of the positive values $R_{j}^{\ast }$ with $j = 1, \dots, N$ is denoted by
$ \overline{R^*} = \frac{1}{N}\sum_{j = 1}^{N} R^*_{j}$.
Values of the bootstrap bivariate bandwidths, estimation errors and relative measures for Models 1 and 2 are included in Table \ref{tab:Bootstrap_S_Bandwidths_RMISE_SBeran}.
Table \ref{tab:Bootstrap_S_Bandwidths_RMISE_SBeran} shows that using bootstrap bandwidth selection increases around $17\%$ to $20\%$ the RMISE error with respect to the optimal bandwidths for Model 1. For Model 2, the RMISE errors when using bootstrap bandwidth selection can be around the double of the optimal ones.

\begin{table}[H]	
	\centering
	\renewcommand{\arraystretch}{1}
	\resizebox{\textwidth}{!}{
		\begin{tabular}{c|c|c|c|c|}
			\cline{2-5}
			& \multicolumn{2}{c|}{Model 1} &   \multicolumn{2}{c|}{Model 2}  \\ \hline
			\multicolumn{1}{|c|}{$P(\delta = 0|X=x)$} & $ 0.2$ & $ 0.5$  & $ 0.2$ & $ 0.5$ \\ \hline
			\multicolumn{1}{|c|}{$h_{MISE}$}   
			& 0.23469 & 0.20408 & 0.11122 & 0.23979    \\
			\multicolumn{1}{|c|}{$g_{MISE}$}   
			& 0.05143 & 0.08102 & 0.91530 & 1.13878    \\
			\multicolumn{1}{|c|}{$RMISE_x(h_{MISE}, g_{MISE})$}  
			& 0.02158 & 0.03024 & 0.05594 & 0.06938    \\ \hline
			\multicolumn{1}{|c|}{ $\overline{h^*} \quad (sd)$ }
			&  0.24172 (0.086) &  0.22600 (0.085)  & 0.31289 (0.132) & 0.36221 (0.142)     \\	
			\multicolumn{1}{|c|}{ $\overline{g^*} \quad (sd)$ }
			& 0.09701 (0.017) & 0.11474 (0.022) & 0.85749 (0.117) &  0.86538 (0.112)     \\
			\multicolumn{1}{|c|}{  $\overline{H^*}$  }   
			& 0.94544 & 0.54256 & 1.82913 &  0.73757      \\ 
			\multicolumn{1}{|c|}{$\overline{RMISE_x(h^*, g^*)}$ } 
			& 0.02769 &  0.03851 & 0.11035 & 0.12633     \\	
			\multicolumn{1}{|c|}{$\overline{R^*}$ }   
			& 0.20027 & 0.17301 & 1.01971 & 0.84987    \\ \hline
			
		\end{tabular}
	}
	\caption{MISE and average bootstrap bandwidths and estimation errors of the smoothed Beran's survival estimator in each level of censoring conditional probability for Models 1 and 2. Numbers within brackets are standar deviations.}
	\label{tab:Bootstrap_S_Bandwidths_RMISE_SBeran}
\end{table}

\begin{figure}[H]
	\centering
	\includegraphics[width=0.45\linewidth]{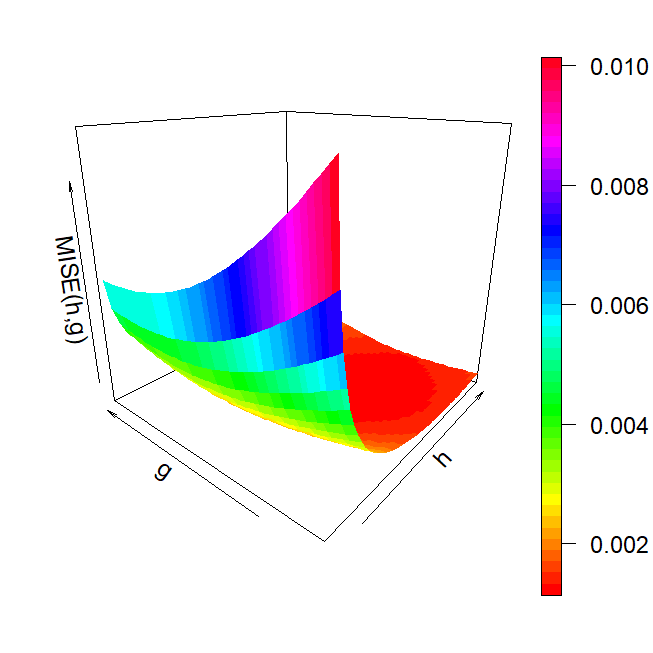}
	\hspace{0.5cm}
	\includegraphics[width=0.45\linewidth]{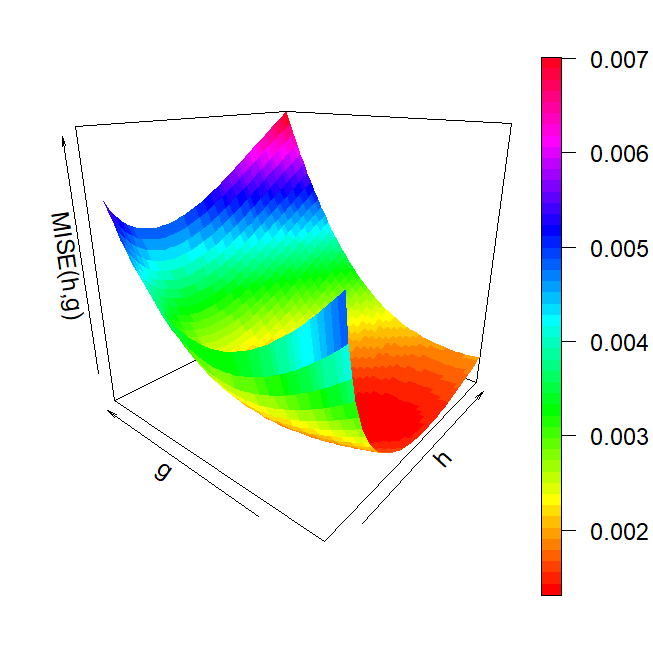}

	
	\caption{$MISE_x(h, g)$ function (left) and $MISE_x^*(h, g)$ function (right) for one sample from Model 1 when $P(\delta = 0|x) = 0.5$.}
	\label{fig:MISE_boot_SBeran}
\end{figure}

Figure \ref{fig:MISE_boot_SBeran} shows the $MISE_x(h, g)$ fuction of the smoothed Beran's estimator and its bootstrap approximation for one sample of Models 1 when the conditional probability of censoring is $0.5$. It is approximated over a meshgrid of $50\times 50$ values of $(h, g)$. Note that both $MISE_x(h, g)$ and $MISE_x^*(h, g)$ curves for each fixed $h$ value are quite similar in the region close to the minimum value of $MISE_x^*(h, g)$. Thus, the influence of the covariate smoothing parameter, $h$, is weak when estimating the survival function using values of the bandwidth $g$ close to the optimal one. The conclusions are similar for Model 2, not included here.

The proposed methods require a high computational cost. Some evidence of this is provided in following paragraphs.

Note that the CPU time of the resampling is the same both when using Beran's estimator and when using the smoothed Beran's estimator. This is due to the fact that the resampling is identical in both cases, except for the perturbation on the life and censoring time variables, which is insignificant in terms of computation time. Regarding the selection of the bootstrap bandwidths, the difference between the two methods lies in the function to be minimised, since it will be unidimensional in the case of Beran's estimator, but two-dimensional in the case of the smoothed Beran's estimator. Since the optimisation of the error function was conducted using a computationally efficient method with no significant differences in the computation times of each scenario, the resulting computation times for both Beran's and the smoothed Beran's estimators are similar. Moreover, the computation time is linear in the number of resamples, $B$.

Table \ref{tab:Bootstrap_S_CPU_times} shows the CPU time required to approximate the bootstrap bandwidth for Beran's estimator by $B = 100$ resamples from one simulated sample of different sizes.
Obtaining the appropriate bootstrap bandwidths to estimate the conditional survival function by 500 resamples from a sample of size 1000 requires 2.5 hours, while a sample of size 3000 requires 2 days. These times seem to increase quadratically as the sample size grows.

\begin{table}[H]
	\centering
	\renewcommand{\arraystretch}{1}
		\begin{tabular}{ccccccc}
			\hline 
			${n}$         & 400 & 1000 & 2000 & 3000 & 6000 & 12000 \\ \hline
			{Time}  &  4.537 &  26.453 & 174.665 & 531.589 & 6715.870 & 27347.460 \\ \hline
			
		\end{tabular}
	\caption{Computation times (in minutes) of the bandwidth selector of Beran's estimator for $N = 1$ sample of size $n$ and $B=100$ bootstrap resamples.}
	\label{tab:Bootstrap_S_CPU_times}
\end{table}

\section{Confidence regions using Beran's and the smoothed Beran's estimators}
\label{Confidence_regions}

Let $x\in I$ be a fixed value of the covariate and consider $S(t|x)$ the conditional survival curve with $t\in I_T$.
The curve $S(t|x)$ belongs to the function space $\mathcal{F}(I_T)$ whose elements are real-valued functions with domain $I_T$. 
From the sample $\{ (X_i, Z_i, \delta_i),$ $i=1, ..., n\}$, Beran's estimation of $S(t|x)$, $\widehat{S}_h(t|x)$, is obtained and a confidence region of $S(t|x)$ at $1 - \alpha$ confidence level associated to Beran's estimator can be constructed. A similar construction is done for the smoothed Beran's estimator. This confidence region of $S(t|x)$ is a random subset of $I_T \times \mathbb{R}$ denoted by $R_{\alpha}$ that satisfies
$$ 
P \big(   (t, S(t|x)) \in R_{\alpha}, \forall  t \in I_T  \big) = 1 - \alpha. 
$$
In this section we propose two different methods to obtain confidence regions of the $S(t|x)$ curve based on resampling techniques. Both Beran's estimator and the smoothed Beran's estimator can be used with these two methods.

\subsection{Method 1 for confidence regions}
\label{Confidence_regions_Method1}

First, Beran's estimator of the conditional survival function, $\widehat{S}_h(t|x)$, given in \eqref{eq:S_Beran} is used. This method follows the ideas of \cite{Cao2010} to obtain prediction regions. It is based on finding the value of $\lambda_{\alpha} \in \mathbb{R}^+$ such that
$$
P\big(|\widehat{S}_h(t|x) -  S(t|x)| <  \lambda_{\alpha}  \sigma(t|x)  , \forall t \in I_T  \big) = 1 - \alpha
$$
with $\sigma^2(t) = Var \big(\widehat{S}_h(t|x)\big)$. Thus, the theoretical confidence region is defined by
$$ 
R_{\alpha}^1 = \Big\{ (t, y): t \in I_T, y \in \big( \widehat{S}_h(t|x) - \lambda_{\alpha}\sigma(t|x),   \widehat{S}_h(t|x) + \lambda_{\alpha}\sigma(t|x)  \big)  \Big\}.
$$

Since $\lambda_{\alpha}$ and $\sigma(t|x)$ are unknown, they are approximated by means of the bootstrap method. The bootstrap confidence region is defined as follows:
$$ 
R_{\alpha}^{1 *} = \Big\{ (t, y): t\in I_T, y \in \big( \widehat{S}_h^*(t|x) - \lambda_{\alpha}^*\sigma^*(t|x),   \widehat{S}_h^*(t|x) + \lambda_{\alpha}^*\sigma^*(t|x)  \big)  \Big\}.
$$
where $ \widehat{S}_h^{*}(t|x)$ is the bootstrap estimation of $S(t|x)$ with bandwidth $h$ and $ \lambda_{\alpha}^*$ and $\sigma^*(t|x)$ are the bootstrap analogue of $\lambda_{\alpha}$ and $\sigma(t|x)$. The confidence region $R_{\alpha}^{1*}$ should satisfy
\begin{equation}
	\label{eq:Bootstrap_S_p_lambda}
	p(\lambda_{\alpha}^*) = P^*\big( (t, \widehat{S}_r(t|x))  \in   R_{\alpha}^{1*}, \forall t\in I_T \big)
	= 1 - \alpha.
\end{equation}

From the original sample $\big\{(X_{i},Z_{i},\delta _{i}) \big\}_{i=1}^{n}$, Beran's estimator of $S(t|x)$ is obtained with appropriate bandwidth $h$, $\widehat{S}_{h}(t|x)$. The algorithm to obtain the bootstrap confidence region for $S(t|x)$ at confidence level $1 - \alpha$ associated to $\widehat{S}_h(t|x)$ is explained below. The Monte Carlo method is used to approximate $\sigma^*(t|x)$, and an iterative method is used to approximate the value of $\lambda_{\alpha}^*$ so that the confidence region has a confidence level approximately equal to $1-\alpha$.

\begin{enumerate}
	\item Compute Beran's estimator $\widehat{S}_r(t|x)$ from the original sample $\big\{(X_i, Z_i, \delta_i)\big\}_{i=1}^{n}$ and pilot bandwidth $r \in I_1$.
	
	\item Generate $B$ bootstrap resamples of the form $\big\{(X_i^{*, k}, Z_i^{*, k}, \delta_i^{*, k})\big\}_{i=1}^{n}$ for $k = 1, \dots, B$ by means of the resampling algorithm for Beran's estimator presented in Subsection \ref{Bandwidth_Beran} and pilot bandwidth $r$.
	
	\item For  $k = 1, \dots, B$, compute $\widehat{S}_h^{*,k}(t|x)$ with the $k$-th bootstrap resample and bandwidth $h$, obtaining $\big\{\widehat{S}_h^{*,k}(t|x)\big\}_{k=1}^B$. 
	
	\item Approximate the standard deviation of $\widehat{S}_h^{*}(t|x)$ by 
	$$
	\sigma^*(t|x) =  \Bigg( \dfrac{1}{B} \sum_{k = 1}^{B}  \bigg( \widehat{S}_h^{*,k}(t|x) -\dfrac{1}{B} \sum_{\text{\textit{l}} = 1}^{B} \widehat{S}_h^{*,\text{\textit{l}}}(t|x) \bigg)^2   \Bigg)^{1/2}, \quad t\in I_T.
	$$
	\item Use an iterative method to obtain an approximation of the value $\lambda_{\alpha}^*$ defined in \eqref{eq:Bootstrap_S_p_lambda}.
	\item The confidence region is given by
	$$
	\widehat{R}_{\alpha}^1= \Big\{ (t, y): t\in I_T, y \in \big( \widehat{S}_h(t|x) - {\lambda}_{\alpha}^*\sigma^*(t|x),   \widehat{S}_h(t|x) + {\lambda}_{\alpha}^*\sigma^*(t|x) \big)   \Big\}.
	$$	
\end{enumerate}

\subsubsection*{Iterative method to approximate $\lambda_{\alpha}^*$} 

The iterative method to approximate the value of $\lambda_{\alpha}^* \in \mathbb{R}^+$, so that the confidence region $R_{\alpha}^*$ has a confidence level approximately equal to $1-\alpha$, is explained below. This algorithm allows to quickly and efficiently approximate the parameter $\lambda_{\alpha}^*$.

Let $\big\{\widehat{S}_h^{*,k}(t|x)\big\}_{k=1}^B$ be the Beran's estimations of the survival function with bandwidth $h$ over a set of $B$ bootstrap resamples of $\big\{(X_i, Z_i, \delta_i)\big\}_{i=1}^{n}$. Define the Monte Carlo approximation of $p(\lambda)$ in \eqref{eq:Bootstrap_S_p_lambda}, for any $\lambda \in \mathbb{R}^+$, as follows:
\begin{equation}
	\label{eq:Bootstrap_S_cobertura_resamples}
	\widehat{p}(\lambda) \simeq \dfrac{1}{B} \sum_{k = 1}^B 
	I \Big(   
	\widehat{S}_r(t|x)  \in 	\big( \widehat{S}_h^{*, k}(t|x) - \lambda\sigma^*(t|x),   \widehat{S}_h^{*, k}(t|x) + \lambda\sigma^*(t|x) \big) , \forall t \in I_T  \Big).
\end{equation}

Let $\lambda_L, \lambda_H  \in \mathbb{R}^+$ be such that 
$  \widehat{p}(\lambda_L) \leq 1 - \alpha \leq \widehat{p}(\lambda_H)$
and let $\zeta > 0$ be a tolerance, for example, $\zeta = 10^{-4}$.

\begin{enumerate}
	\item Obtain $\lambda_M = \dfrac{\lambda_L + \lambda_H}{2}$ and compute Monte Carlo approximations of  $\widehat{p}(\lambda_L)$, $\widehat{p}(\lambda_M)$ and $\widehat{p}(\lambda_H)$ according to \eqref{eq:Bootstrap_S_cobertura_resamples}. 
	
	\item  If $\widehat{p}(\lambda_M) = 1- \alpha$ or $\widehat{p}(\lambda_H) - \widehat{p}(\lambda_L) < \zeta$, then  $\lambda_{\alpha}^{*} = \lambda_M$. Otherwise,
	
	\begin{enumerate}
		\item If $1 - \alpha < \widehat{p}(\lambda_M)$, then $\lambda_H = \lambda_M$ and return to Step 1.
		\item If $\widehat{p}(\lambda_M) < 1 - \alpha $, then $\lambda_L = \lambda_M$ and return to Step 1.
	\end{enumerate}	
\end{enumerate}

This method to obtain confidence regions for the curve $S(t|x)$ for fixed $x \in I$ and $t$ covering $I_T$ based on Beran's estimator can be adapted to obtain confidence regions using the smoothed Beran's estimator. Simply replace Beran's estimator, $\widehat{S}_{h}(t|x)$, by the smoothed Beran's estimator, $\widetilde{S}_{h,g}(t|x)$, given in \eqref{eq:S_smoothed_estimator} where necessary, and obtain the analogous bootstrap approximations of ${\lambda}_{\alpha}$ and $\sigma(t|x)$:

\begin{enumerate}
	\item Compute the smoothed Beran's estimator  $\widetilde{S}_{r, s}(t|x)$ from the original sample $\big\{(X_i, Z_i, \delta_i)\big\}_{i=1}^{n}$ and pilot bandwidths $r \in I_1$ and $s\in I_2$.
	
	\item Generate $B$ bootstrap resamples of the form $\big\{(X_i^{*, k}, Z_i^{*, k}, \delta_i^{*, k})\big\}_{i=1}^{n}$ for $k = 1, \dots, B$ by means of the resampling algorithm for the smoothed Beran's estimator presented in Subsection \ref{Bandwidth_SBeran} and pilot bandwidths $r$ and $s$.
	
	\item For  $k = 1, \dots, B$, compute $\widetilde{S}_{h, g}^{*,k}(t|x)$ with the $k$-th bootstrap resample and bandwidths $h$ and $g$, obtaining $\big\{\widetilde{S}_{h, g}^{*,k}(t|x)\big\}_{k=1}^B$. 
	
	\item Approximate the standard deviation of $\widetilde{S}_{h, g}^{*}(t|x)$ by 
	$$
	\sigma^*(t|x) = \Bigg( \dfrac{1}{B} \sum_{k = 1}^{B}  \bigg( \widetilde{S}_{h, g}^{*,k}(t|x) -\dfrac{1}{B} \sum_{\text{\textit{l}} = 1}^{B} \widetilde{S}_{h, g}^{*,\text{\textit{l}}}(t|x) \bigg)^2   \Bigg)^{1/2}, \quad t\in I_T.
	$$
	\item Use an iterative method to obtain an approximation of the value $\lambda_{\alpha}^*$ defined in \eqref{eq:Bootstrap_S_p_lambda}.
	\item The confidence region is given by
	$$
	\widetilde{R}_{\alpha}^1= \Big\{ (t, y) : t \in I_T, y \in \big( \widetilde{S}_{h, g}(t|x) - {\lambda}_{\alpha}^*\sigma^*(t|x),   \widetilde{S}_{h, g}(t|x) + {\lambda}_{\alpha}^*\sigma^*(t|x) \big)    \Big\}.
	$$	
\end{enumerate}

The pilot bandwidths defined in \eqref{eq:Bootstrap_S_piloto_r} and \eqref{eq:Bootstrap_S_piloto_s} are used for the confidence region algorithm based on both Beran's and the smoothed Beran's estimators.

\subsection{Method 2 for confidence regions}
\label{Confidence_regions_Method2}

An alternative procedure to obtain a confidence region for $S(t|x)$, with fixed $x \in I$ and $t$ covering the interval $I_T$, is based on considering that the curve $S(t|x)$ belongs to the functional space $L_p(I_T)$ defined using the $p$-norm for functional vector spaces, $\|\cdot\|_p$. 
The confidence region for $S(t|x)$  computed at the $1 - \alpha$ confidence level is a ball around $\widehat{S}_{h}(t|x)$ of radius $\rho_{\alpha}$, where $\rho_{\alpha}$ is such that
$$
P\big(  \| \widehat{S}_{h}(t|x) - S(t|x) \|_{p} < \rho_{\alpha}  \big) = 1 - \alpha.
$$
This idea was presented in \cite{Zhun2017} to obtain prediction regions in functional autoregression models. Since $S(t|x)$ is unknown, the distribution of $Q =  \| \widehat{S}_{h}(t|x) - S(t|x) \|_{p} $ is not available and the value of $\rho_{\alpha}$ can not be calculated. Therefore, a bootstrap approximation is given by $Q^* =  \| \widehat{S}_{h}^*(t|x) - \widehat{S}_r(t|x) \|_{p} $. The bootstrap confidence region is a ball in $L_p(I_T)$ around $\widehat{S}_{h}(t|x)$ of radius $\rho_{\alpha}^*$, where  $\rho_{\alpha}^*$ is such that
$$
P^*\big(   \|\widehat{S}_{h}^{*}(t|x) -\widehat{S}_{r}(t|x) \|_{p} < \rho_{\alpha}^*    \big) = 1 - \alpha.
$$
and $r \in I_1 $ is an auxiliary bandwidth.

From the original sample $\big\{(X_{i},Z_{i},\delta _{i}) \big\}_{i=1}^{n}$, Beran's estimator of $S(t|x)$ is obtained with appropriate bandwidth $h$, $\widehat{S}_{h}(t|x)$. The algorithm to obtain the bootstrap confidence region for $S(t|x)$ at confidence level $1 - \alpha$ associated to $\widehat{S}_h(t|x)$ is explained below. The Monte Carlo method is used to approximate the radius $\rho_{\alpha}$, so that the confidence region has a confidence level approximately equal to $1-\alpha$.

\begin{enumerate}
	\item Compute Beran's estimator $\widehat{S}_{r}(t|x)$ with the original sample $\{(X_i, Z_i, \delta_i)\}_{i=1}^n$ and pilot bandwidth $r \in I_h$.
	
	\item Generate $B$ bootstrap resamples  $\big\{(X_i^{*k}, Z_i^{*k}, \delta_i^{*k})\big\}_{i=1}^n$, for $k = 1, ..., B$, by means of the resampling algorithm for Beran's estimator presented in Subsection \ref{Bandwidth_Beran} and pilot bandwidth $r$.
	
	\item For $k = 1, ..., B$, compute $\widehat{S}_{h}^{*k}(t|x)$ with the $k$-th bootstrap resample and bandwidth $r$ and obtain 
	$$ 
	Q^*_{k} =  \| \widehat{S}_{h}^{*k}(t|x) - \widehat{S}_r(t|x) \|_{p} 
	$$
	
	\item Sort the values $Q^*_{1}, ..., Q^*_{B}$ by obtaining $Q^*_{(1)}, ..., Q^*_{(B)}$ and select $\rho^*_{\alpha} = Q^*_{([B(1 - \alpha)])}$.
	
	\item The confidence region is the ball in $L_p(I_T)$ around $\widehat{S}_{h}(t|x)$ with radius $\rho_{\alpha}^*$.
	
\end{enumerate}

Regarding the norm to be used, the usual norms of the function spaces $L_1$ and $L_2$ allow us to mathematically define the confidence region and to check whether or not a given curve belongs to this region. The disadvantage of these function spaces is that they do not allow a graphical representation of the confidence region. 

Choosing the function space $L_{\infty}$ and its associated norm, $\|\cdot\|_{\infty}$, then the statistic used to obtain the confidence region is defined as follows
$$ 
Q = \| \widehat{S}_{h}(t|x) - S(t|x) \|_{\infty} =  \sup_{t \in I_T} | \widehat{S}_{h}(t|x) - S(t|x) |
$$
and the confidence region is
$$ 
R_{\alpha}^2 = \Big\{ (t, y) : t\in I_T, y \in \big( \widehat{S}_h(t|x) - \rho_{\alpha},   \widehat{S}_h(t|x) + \rho_{\alpha} \big)   \Big\}.
$$
whose graphical representation may be useful. The disadvantage of this choice of the space is that the confidence region $R_{\alpha}$ has the same radius $\rho_{\alpha}$ at all points $t \in I_T$, so it does not capture the variability of the estimator, $\sigma^2(t) = Var (\widehat{S}_h(t|x)) $.

This method can be adapted to obtain confidence regions using the smoothed Beran's estimator. Simply replace Beran's survival estimator $\widehat{S}_{h}(t|x)$ by the smoothed Beran's estimator  $\widetilde{S}_{h,g}(t|x)$ given in \eqref{eq:S_smoothed_estimator} where necessary.
The confidence region for $S(t|x)$ based on the smoothed Beran's estimator at $1 - \alpha$ confidence level is a ball in $L_p(I_T)$ around $\widetilde{S}_{h, g}(t|x)$ of radius $\rho_{\alpha}$, where $\rho_{\alpha}$ is such that
$$
P(Q < \rho_{1 - \alpha}) = 1 - \alpha
$$
with
$$
Q = \| \widetilde{S}_{h, g}(t|x) - S(t|x) \|_{p}.
$$
A similar procedure to the one shown in the previous paragraphs for Beran's estimator allows us to obtain the bootstrap approximation of $\rho_{\alpha}$. 

The pilot bandwidths defined in \eqref{eq:Bootstrap_S_piloto_r} and \eqref{eq:Bootstrap_S_piloto_s} are used for the confidence region algorithm based on both Beran's and the smoothed Beran's estimators. 

If the aim were the point estimation of $S(t|x)$, the algorithms proposed in this section can be easily adapted to obtain confidence intervals for $S(t|x)$ for fixed $x \in I$ and $t \in I_T$.

\section{Simulation study for confidence regions}
\label{Confidence_Simulation}

A simulation study is carried out to analyse the performance of the bootstrap confidence regions obtained by means of the two methods proposed in Section \ref{Confidence_regions} and based on both Beran's and the smoothed Beran's estimator.

Models 1 and 2 are considered and the simulation setup is the one introduced in Section \ref{Bandwidth_Simulation}. The number of simulated samples of each model is $N = 300$ and $B = 500$ bootstrap resamples are obtained for each sample. The sample size is $n = 400$. The confidence level is $1 - \alpha$ with $\alpha = 0.05$.
When Beran's estimator is considered, the optimal bandwidth that minimises the mean integrated squared error is used ($h = h_{MISE}$ from Table \ref{tab:Bootstrap_S_Bandwidths_RMISE_Beran}). Similarly, the two-dimensional bandwidth that minimises the MISE is considered when using the smoothed Beran's estimator ($(h, g) = (h_{MISE}, g_{MISE})$ from Table \ref{tab:Bootstrap_S_Bandwidths_RMISE_SBeran}). These bandwidths are unknown in practice, but they allow a fair comparison of the methods in the simulation study. 
Regarding the pilot bandwidth defined in \eqref{eq:Bootstrap_S_piloto_r}, Model 1 considers $c = 3/2$, while Model 2 considers $c = 1$, as explained in Section \ref{Bandwidth_selection}.

Denote the lower and upper bounds of the confidence region by $l(t, x)$ and $u(t, x)$, respectively. It may happen that the lower bound of the confidence region is less than 0 or the upper bound is greater than one for some points $(t_0, x_0)$. When this happens, we set $l(t_0, x_0) = 0$ or $u(t_0, x_0) = 1$, as appropriate.

It is clear that $S(t|x) = 1$ when $t = 0$ and $S(t|x)$ is not necessarily 1 when $t = 0 + \varepsilon$ with any $\varepsilon > 0$. However, due to the lack of information provided by the data at times close to zero, it is the case that the estimation of $S(t|x)$ is 1 for the smallest values of the time grid in most of the samples of the study. As a consequence, using Method 1, $l(t, x) = 1 = u(t, x)$ for such small values of $t$ and the confidence region does not contain the true survival curve, so the coverage decreases. The proposed solution is to artificially increase the width of the confidence region at the first points of the grid: for those values of $t$ such that $l(t, x) = 1 = u(t, x)$, we make $l(t, x) = l(t^{\prime}, x)$ where $t^{\prime} \in \{t_1, \dots, t_n\}$ is the first grid point such that $l(t_0, x) < 1$. This is a problem that method 2 does not present, since the width of the confidence region is constant at all points of the time grid and the variability of the conditional survival estimations in the right tail of the time distribution causes this width to increase.

A confidence region performs well if its coverage is close to the nominal one, in this case $1 - \alpha = 0.95$, and has a small area or average width. The following values measure the performance of the confidence region and allow for the comparison of results.

Coverage is the percentage of bootstrap regions that contain the whole theoretical survival curve and it is defined as follows
\begin{equation*}
	\label{eq:S_coverage}
	\dfrac{1}{N} \sum_{j = 1}^N I\Big\{  S(t_k|x) \in \big( l(t_k, x) , u(t_k, x) \big), \forall k = 1, ..., n_T \Big\}.
\end{equation*}
The mean pointwise coverage is the mean of the proportion of time grid values for which the confidence region contains the theoretical conditional survival curve. It is given by
\begin{equation*}
	\label{eq:S_pointwise_coverage}
	\dfrac{1}{N} \sum_{j = 1}^N \bigg(   \dfrac{1}{n_T}  \sum_{k=1}^{n_T}  I\Big\{   S(t_k|x) \in \big( l(t_k, x) , u(t_k, x) \big) \Big\}   \bigg).
\end{equation*}
The average width of the bootstrap confidence region is defined by
\begin{equation*}
	\label{eq:S_width}
	\dfrac{1}{N} \sum_{j = 1}^N \Bigg(   \dfrac{1}{n_T}  \sum_{k=1}^{n_T} \big( u(t_k, x) - l(t_k, x)   \big) \Bigg).
\end{equation*}

Winkler score (see \cite{Winkler1972}) is also used to compare the behaviour of the methods. For classical confidence or prediction intervals, it is defined as the length of the interval plus a penalty if the theoretical value is outside the interval. Thus, it combines width and coverage. For values that fall within the interval, the Winkler score is simply the length of the interval. So low scores are associated with narrow intervals. When the theoretical value falls outside the interval, the penalty is proportional to how far the observation is from the interval. The formula of the Winkler score (WS) as a function of the time and covariate variables is as follows:
\begin{equation*}
	\begin{array}{rcl}
		\text{WS} (t, x)  & = & u(t, x) - l(t, x) + \dfrac{2}{\alpha} (l(t,x) - S(t|x)) I\big(  S(t|x) < l(t, x)  \big) 
		\\
		& & + \dfrac{2}{\alpha} (S(t|x) - u(t, x)) I\big(  S(t|x) > u(t, x)  \big).
	\end{array}
	\label{eq:S_WS}
\end{equation*}
Since we are working with confidence regions for fixed $x\in I$ and $t$ varying over the interval $I_T$, the integrated Winkle score is proposed as a criteria for the comparison of the confidence regions. It is defined by
\begin{equation*}
	IWS(x) = \int_{I_T} WS(t, x) dt.
	\label{eq:S_IWS}
\end{equation*}
and the lower the value of IWS, the better the performance of the confidence region. 

The results obtained are shown in Tables \ref{tab:Bootstrap_S_confidence_region_Mod2} and \ref{tab:Bootstrap_S_confidence_region_Mod3}.
The high values of pointwise coverage in all scenarios are remarkable. Furthermore, these coverage percentages are preserved when using double smoothing, while the average width of the confidence regions is decreased.
This is reflected in the IWS, which presents much larger values in the Beran's estimator-based confidence regions.

Method 1 has lower mean coverage, but equal pointwise coverage and smaller width than Method 2 in Model 1. In Model 2, the coverages of the two methods are similar, with Method 1 providing confidence regions of smaller width. The coverage indicates the percentage of times the theoretical curve is completely contained in the confidence band. This coverage decreases as soon as the curve goes outside the region at a single point on the time grid. This, that only a few points go out of the region, is what mainly happens here.

In some of the scenarios, the mean coverage of Method 1 is remarkably low. For example, the average coverage of the confidence region based on the Beran's estimator for Model 1 is $40\%$. This value indicates that only in 60 out of 100 trials does the confidence region obtained by the proposed method entirely contain the theoretical curve. However, in the same scenario, the average point coverage is $96\%$, so the survival curve is within the confidence region at 96 out of 100 grid points, which is a considerably high value of the pointwise coverage.

In conclusion, the two proposed methods for the confidence regions have reasonable behaviours, both presenting very high pointwise coverages. Method 1 provides confidence regions of variable width at the cost of slightly decreasing the average coverage. Method 2 has higher coverage percentages but also a larger width, which is also constant everywhere. The results obtained using the smoothed Beran's estimator in either method are promising.

\begin{table}[H]
	\centering
	\renewcommand{\arraystretch}{1}
	\resizebox{\textwidth}{!}{
		\begin{tabular}{|c|c|c|c|c|c|c|c|c|}
			\hline
			{Model 1} & \multicolumn{4}{|c|}{Beran} & 
			\multicolumn{4}{|c|}{SBeran} \\ \hline
			$P(\delta =0 \mid X = 0.6)$ & \multicolumn{2}{|c|}{$0.2$} & \multicolumn{2}{|c|}{$0.5$} & 
			\multicolumn{2}{|c|}{$0.2$} & \multicolumn{2}{|c|}{$0.5$} \\ \hline
			Method  & Meth 1 & Meth 2 & Meth 1 & Meth 2 & Meth 1 & Meth 2 & Meth 1 & 
			Meth 2 \\ \hline
			Width 
			& 0.16264 & 0.21677 & 0.21664 & 0.35643 & 0.15759 & 0.16426 & 0.17985 &  0.21985 \\ \hline
			Coverage ($\%$) 
			& 39.33 & 97.67 & 40.33 & 97.00 & 97.67 & 97.33 & 58.67 & 98.67 \\ \hline
			$%
			\begin{array}{c}
				\text{Pointwise} \\ 
				\text{coverage(}\%\text{)}%
			\end{array}%
			$ 
			& 96.45 & 99.93 & 95.85 & 99.82 & 98.71 & 99.44 & 96.57 & 99.67 \\ \hline
			IWS
			& 0.15076 & 0.17167 & 0.21480 & 0.26943 & 0.13759 & 0.13550 & 0.16372 & 0.17469 \\ \hline
		\end{tabular}
	}	
	\caption{Coverage, average width and IWS of the $95\%$ confidence regions by means Methods 1 and 2 and Beran's and the smoothed Beran's estimators using $N = 300$ simulated samples from Model 1.}
	\label{tab:Bootstrap_S_confidence_region_Mod2}
\end{table}

\begin{table}[H]
	\centering
	\renewcommand{\arraystretch}{1}
	\resizebox{\textwidth}{!}{
		\begin{tabular}{|c|c|c|c|c|c|c|c|c|}
			\hline
			{Model 2} & \multicolumn{4}{|c|}{Beran} & 
			\multicolumn{4}{|c|}{SBeran} \\ \hline
			$P(\delta =0 \mid X = 0.8)$ & \multicolumn{2}{|c|}{$0.2$} & \multicolumn{2}{|c|}{$0.5$} & 
			\multicolumn{2}{|c|}{$0.2$} & \multicolumn{2}{|c|}{$0.5$} \\ \hline
			Method  & Meth 1 & Meth 2 & Meth 1 & Meth 2 & Meth 1 & Meth 2 & Meth 1 & 
			Meth 2 \\ \hline
			Width 
			& 0.34203 & 0.34511 & 0.42486 & 0.41146 & 0.24070 & 0.19981 & 0.37740 & 0.27440 \\ \hline
			Coverage ($\%$) 
			& 85.33 & 89.00 & 66.67 & 83.33 & 88.67 & 93.67 & 96.00 & 99.67 \\ \hline
			$%
			\begin{array}{c}
				\text{Pointwise} \\ 
				\text{coverage(}\%\text{)}%
			\end{array}%
			$
			& 97.56 & 99.32 & 92.90 & 98.91 & 98.24 & 98.94 & 98.67 & 99.94 \\ \hline
			IWS 
			& 1.37220 & 1.18335 & 2.06192 & 1.38213 & 0.93535 & 0.77965 & 1.45099 & 0.92742 \\ \hline
		\end{tabular}
	}	
	\caption{Coverage, average width and IWS of the $95\%$ confidence regions by means Methods 1 and 2 and Beran's and the smoothed Beran's estimators using $N = 300$ simulated samples from Model 2.}
	\label{tab:Bootstrap_S_confidence_region_Mod3}
\end{table}

\begin{figure}[H]
	\centering
	\includegraphics[width=0.495\linewidth]{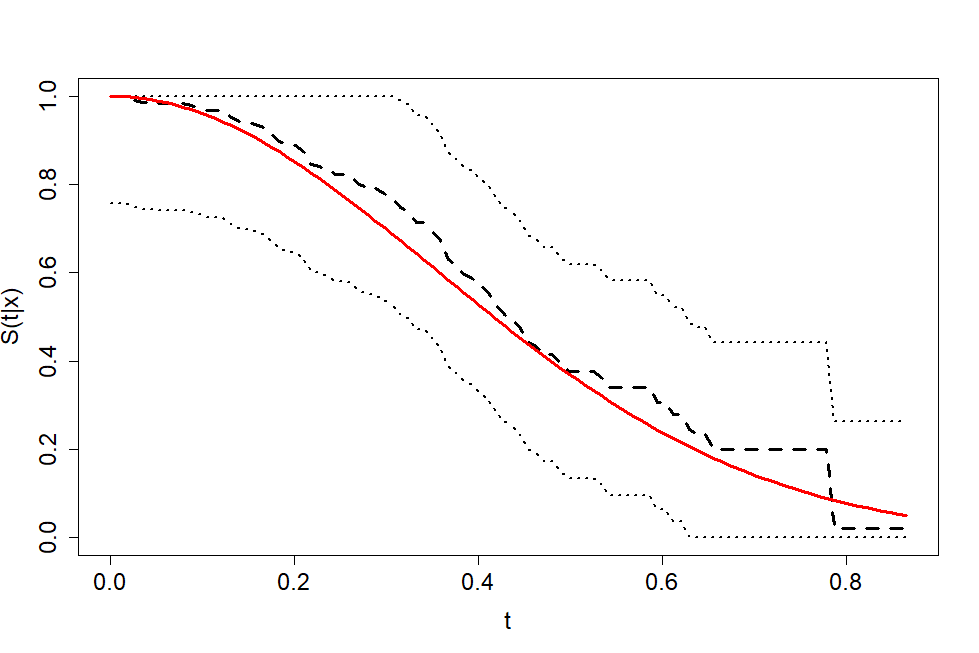}
	\includegraphics[width=0.495\linewidth]{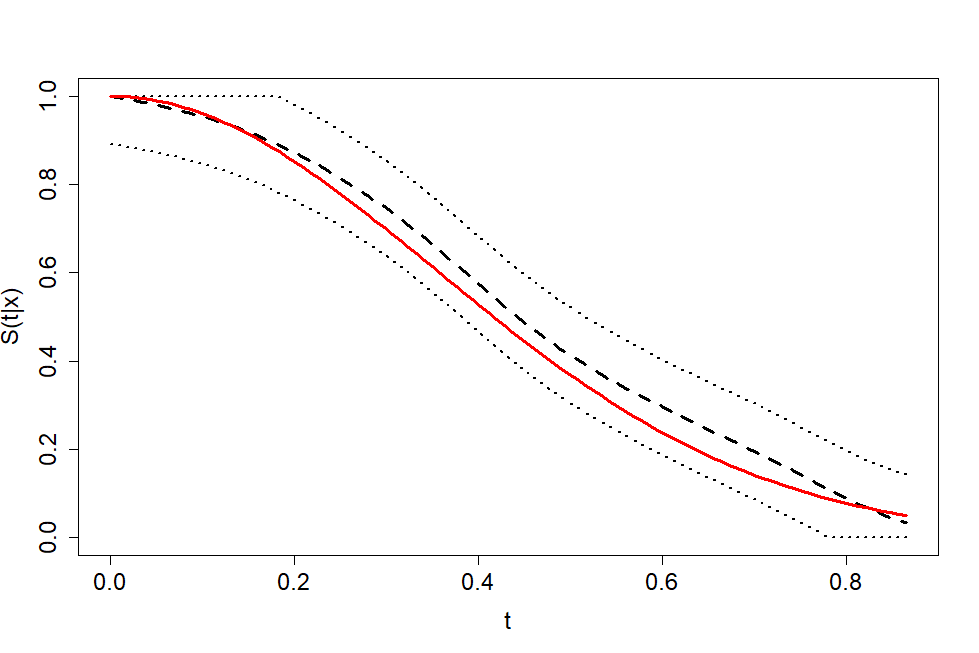}
	
	\includegraphics[width=0.495\linewidth]{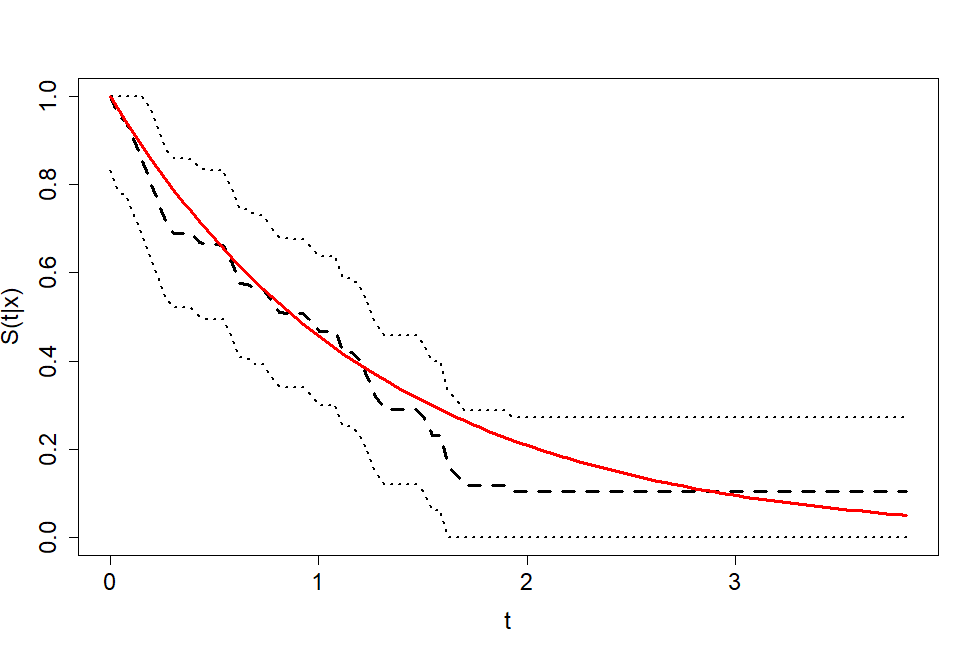}
	\includegraphics[width=0.495\linewidth]{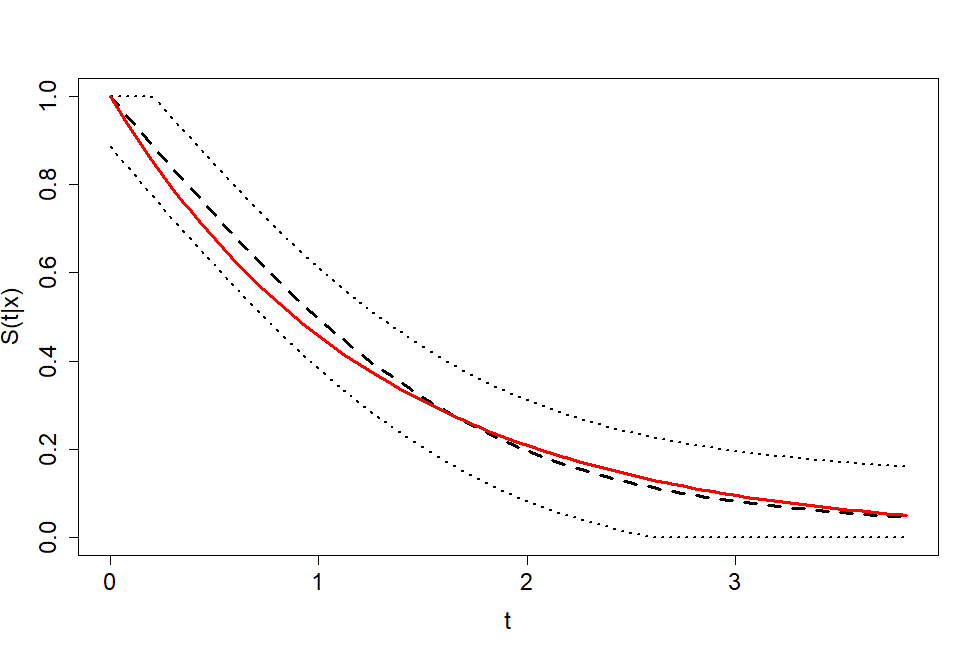}
	\caption{Theoretical $S(t|x)$ (red solid line), estimation with MISE bandwidths (black dashed line) and $95\%$ confidence region (black dotted lines) by means of Beran's estimator (left) and the smoothed Beran's estimator (right) for one sample from Model 1 (top) and Model 2 (bottom) when $P(\delta = 0 | x) = 0.5$ using Method 2.}
	\label{fig:Bootstrap_S_band_Beran_SBeran_Method2}
\end{figure}

This analysis is also illustrated in following figures. 
Figure \ref{fig:Bootstrap_S_band_Beran_SBeran_Method1} shows the confidence regions for the conditional survival function obtained by Method 1 for one sample from Models 1 and 2. The confidence regions obtained by Method 2 are shown in Figure \ref{fig:Bootstrap_S_band_Beran_SBeran_Method2}. The higher variability of the Beran's estimations in the resamples with respect to the smoothed Beran's estimations leads to much wider confidence regions. When using Method 1, this only affects the width of the confidence region at the right tail of the time distribution. When using Method 2, this variability causes the confidence region to have a larger width for all points on the time grid.

\begin{figure}[H]
	\centering
	
	\includegraphics[width=0.495\linewidth]{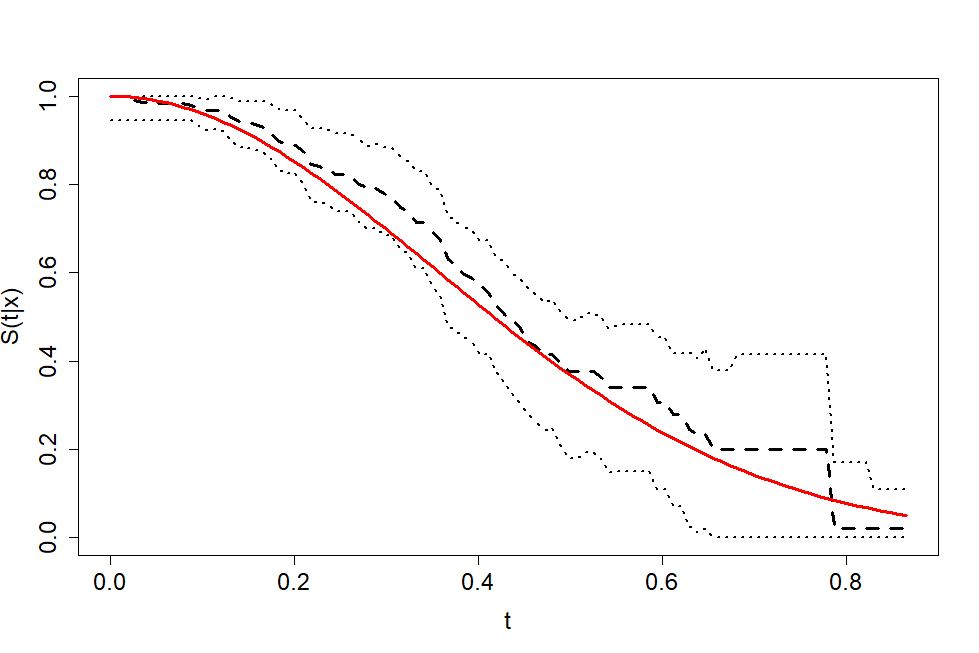}
	\includegraphics[width=0.495\linewidth]{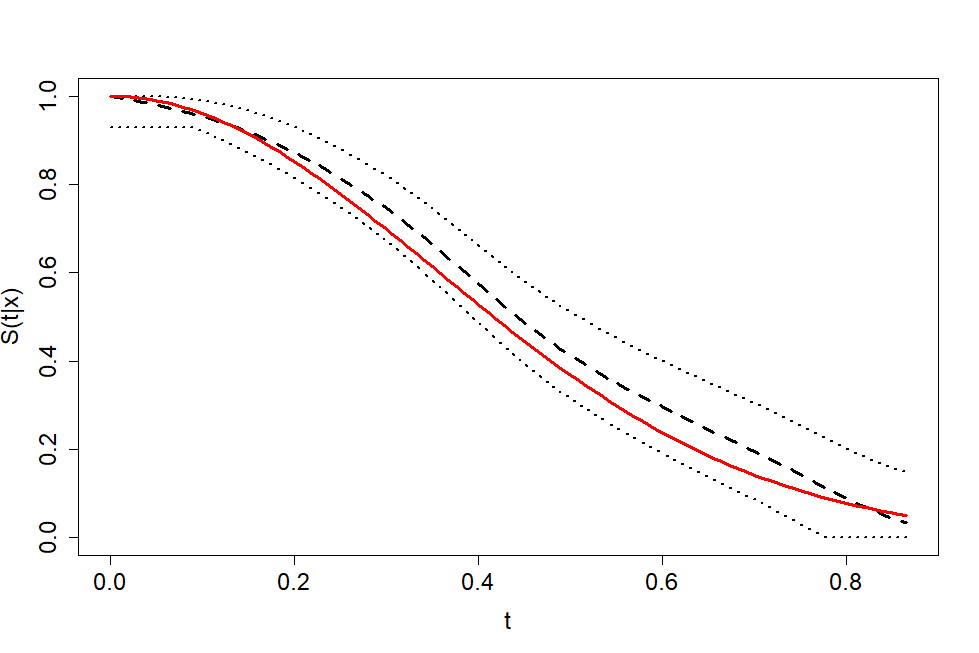}
	
	\includegraphics[width=0.495\linewidth]{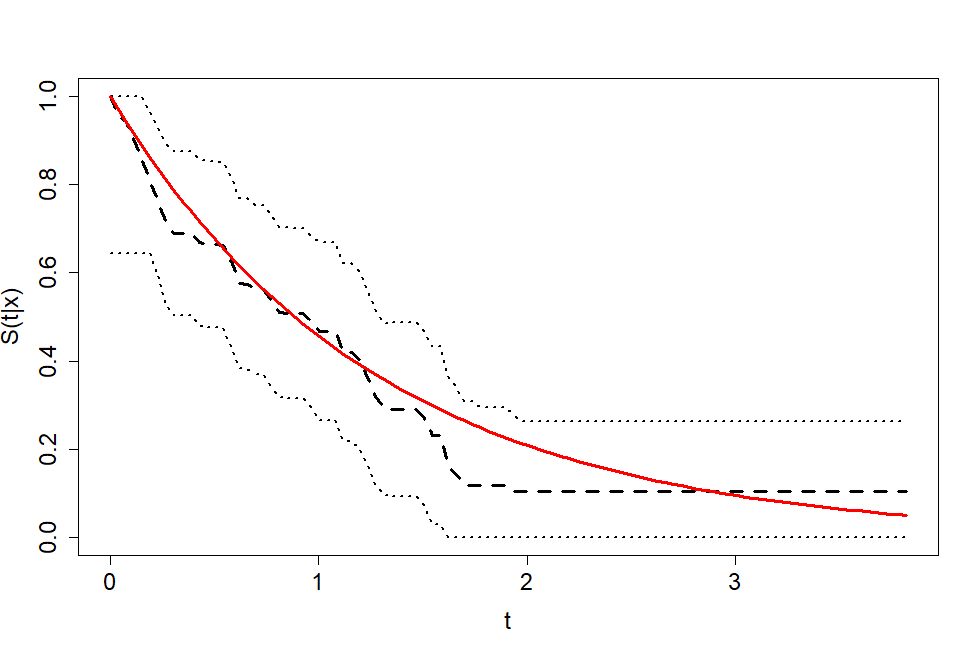}
	\includegraphics[width=0.495\linewidth]{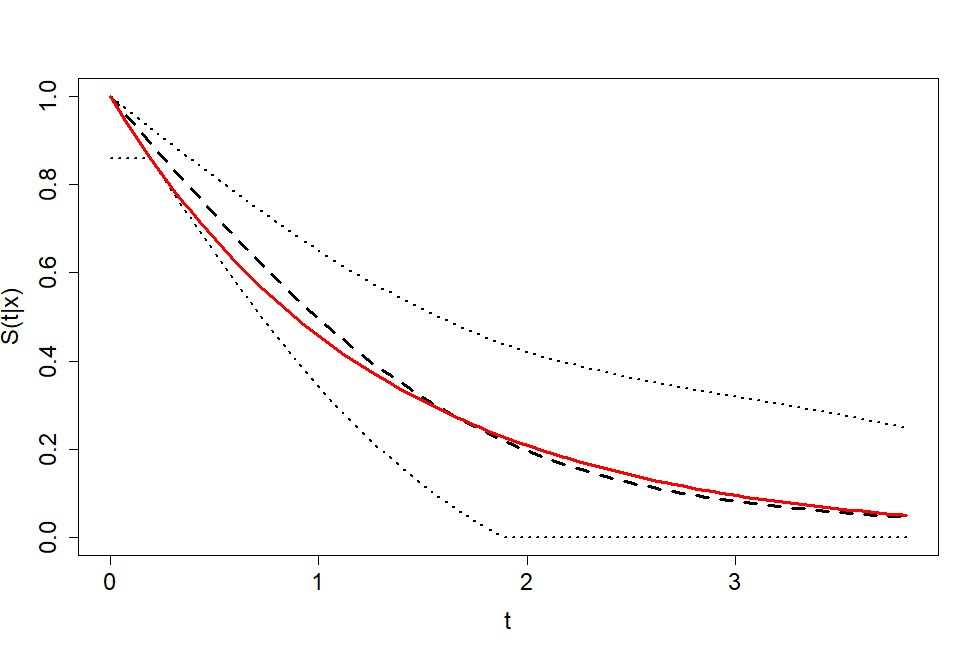}
	
	\caption{Theoretical $S(t|x)$ (red solid line), estimation with MISE bandwidths (black dashed line) and $95\%$ confidence region (black dotted lines) by means of Beran's estimator (left) and the smoothed Beran's estimator (right) for one sample from Model 1 (top) and Model 2 (bottom) when $P(\delta = 0 | x) = 0.5$ using Method 1.}
	\label{fig:Bootstrap_S_band_Beran_SBeran_Method1}
\end{figure}

The computation times required to compute the confidence regions based on both methods are similar. In fact, the times are comparable to those shown in Section \ref{Bandwidth_Simulation}, since the really slow part of all these procedures is obtaining the bootstrap resamples.

\section{Estimation of the conditional survival function for length-of-stay of hospital ward and ICU for COVID-19 patients}
\label{Real_data}

The usefulness of the automatic bootstrap selector of the bandwidths of Beran's and the smoothed Beran's estimator is illustrated in this section.
The survival function of the time that COVID-19 patients remain hospitalised in ward or the Intensive Care Unit (ICU) is estimated by means of Beran's and the smoothed Beran's estimators. A dataset of $n = 2484$ patients from SERGAS (Galician health service) with dates of admission and discharge (if applicable), age, sex and previous diseases of COVID-19 patients in Galicia (Spain) has been used. The data corresponds to patients suffering from COVID-19 being admitted in the public hospital system in Galicia since the beginnng of the pandemic until May 11th 2020.

The event of interest is the patient leaving ward, so the time variable which is subject to right random censoring is the length-of-stay in hospital ward. An informative covariate of the survival time is the age of the patient. Other factors like sex or previous diaseases are used to disaggregate interesting subpopulations. There are certain risk factors for COVID-19 that could affect hospitalisation and recovery times. Two of these are obesity and COPD. COPD (chronic obstructive pulmonary disease) is a chronic inflammatory lung disease that causes obstructed airflow from the lungs. The following subsections consider whether or not patients have obesity or COPD in order to analyse their influence on the hospitalisation times.

\subsection{Length-of-stay in hospital ward}

The time until a COVID-19 positive patient leaves the ward is first considered. A patient leaves the ward because he/she is discharged, admitted to the ICU or dies. When none of these three circumstances is observed for a patient before the end of the study, the censoring time is observed. The total number of hospitalised patients followed up is 2453 and the censoring percentage of this dataset is $8.85\%$.

\begin{figure}[H]
	\centering
	\includegraphics[width=0.45\linewidth]{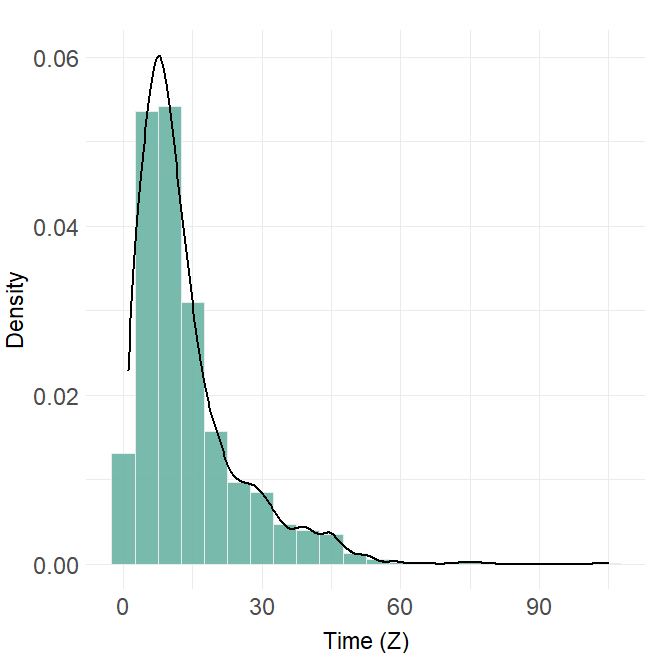}
	\includegraphics[width=0.45\linewidth]{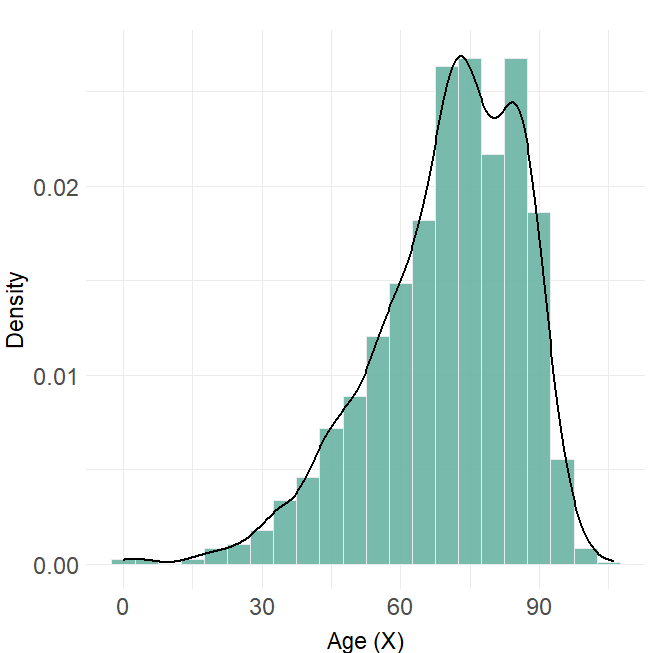}
	\caption{Histogram and kernel density estimation for time in ward (left) and age (right).}
	\label{fig:Bootstrap_S_Real_data_Time_WARD_histogram}
\end{figure}

Table \ref{tab:Bootstrap_S_Real_data_summary_WARD} shows summary statistics of the hospitalisation time in ward and the age of COVID-19 patients disaggregating the censored and uncensored groups. Figure \ref{fig:Bootstrap_S_Real_data_Time_WARD_histogram} shows the histogram  and the kernel density estimation of the time in ward and the age for all patients.

\begin{table}[H]
	\centering
	\renewcommand{\arraystretch}{1}
	\begin{tabular}{ccrrrrrr}
		\hline
		& & {min.} & ${1^{st} Q.}$ & {median} & {mean} & ${3^{th} Q.}$ & {max.} \\ \noalign{\hrule height 1pt}  
		\multirow{1}{*}{{Censored data}} & Time & 1.00 & 5.00 & 15.00 & 18.22 & 28.00 & 105.00 \\
		& Age & 4.00 & 69.00 & 80.00 & 76.04 & 87.00 & 100.00 \\ 
		\multirow{1}{*}{{Uncensored data}} & Time & 1.00 & 6.00 & 10.00  & 13.02 & 16.00 & 75.00  \\
		& Age & 0.00 & 60.00  & 72.00 & 69.61  & 82.00 & 106.00 
		\\ \hline
	\end{tabular}
	\caption{Summary statistics for length-of-stay in ward $(Z)$ and age $(X)$ for the uncensored group (patients who left ward) and the censored group (patients in ward).}
	\label{tab:Bootstrap_S_Real_data_summary_WARD}
\end{table}

The bootstrap algorithm shown in Subsection \ref{Bandwidth_SBeran} is used here to compute the bootstrap bandwidths for estimating the survival function of the time in ward of the Galician COVID-19 patients. Due to the good results that the smoothed Beran's estimator showed in the previous simulations, this is the estimator mainly used in this section. Some interesting confidence regions based on the smoothed Beran's estimator are also obtained. Method 1, which was proposed in Subsection \ref{Confidence_regions_Method1} and provides confidence regions of varying width, is used.

The bootstrap estimation is obtained in a grid of time $t_1 < \cdots < t_{n_T}$ with $t_{n_T} = \widehat{Q}(0.95)$ and $n_T = 100$. The pilot bandwidth for the covariate used in the bootstrap algorithm is the one defined in \eqref{eq:Bootstrap_S_piloto_r} with $c = 3/2$. The pilot bandwidth for the time variable is the one defined in \eqref{eq:Bootstrap_S_piloto_s}.
Three age profiles are considered: 40, 60 and 80 years old. In some cases, because of sample limitations, only 60 and 80 year old profiles will be considered.

The bandwidth that minimises the Monte Carlo approximation of the bootstrap MISE for Beran's estimator, $MISE^*_x(h^*)$, and the two-dimensional bandwidth that minimises the Monte Carlo approximation of the bootstrap MISE for the smoothed Beran's estimator, $MISE^*_x(h^*, g^*)$, are shown in Table \ref{tab:Bootstrap_S_Time_WARD_bootstrap_bandwidths}.
For $x = 80$, the $RMISE^*_x(h, g)$ function is decreasing in $h$, so the bandwidth selector for the smoothed Beran's estimator proposes the upper end of the interval considered for this smoothing parameter as the bootstrap bandwidth.

\begin{table}[H]
	\begin{center}
		\renewcommand{\arraystretch}{1}
		\begin{tabular}{cccc}
			\hline
			& \multicolumn{1}{c}{Beran} & \multicolumn{2}{c}{SBeran} \\ \noalign{\hrule height 1pt}  
			\hspace*{0.2cm} $x$ \hspace*{0.2cm}  & \hspace*{0.2cm} $ h^*$ \hspace*{0.2cm}  & $h_2^*$ & $g_2^*$  \\
			
			40 & 4.765306  & 5.507370 & 1.266695  \\ 
			
			60 & 4.571429  & 5.548651 & 0.784956  \\ 
			
			80 & 13.387760 & 30.000000 & 2.348188  \\ 
			\hline 
		\end{tabular}
		\caption{Bootstrap bandwidth for Beran's estimation and the smoothed Beran's estimation of the conditional survival function of the time in ward for some different values of age.}
		\label{tab:Bootstrap_S_Time_WARD_bootstrap_bandwidths}	
	\end{center}
\end{table}

\begin{figure}[H]
	\centering
	\includegraphics[width=0.45\linewidth]{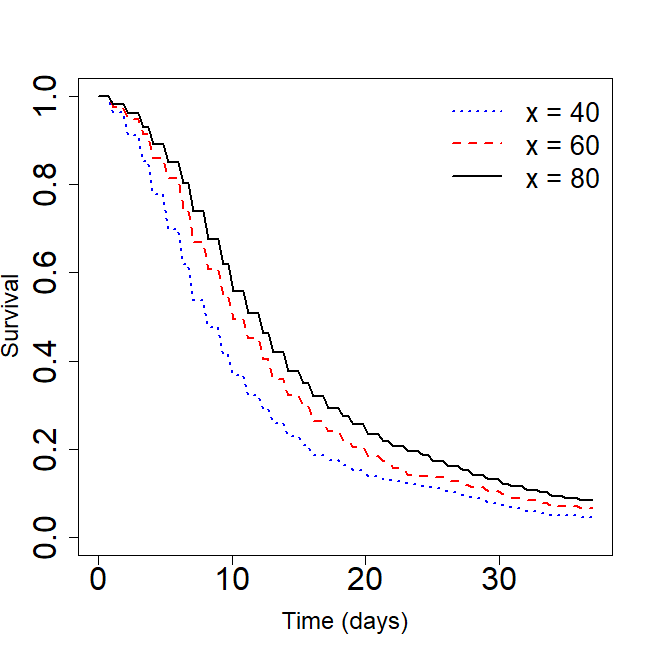}
	\includegraphics[width=0.45\linewidth]{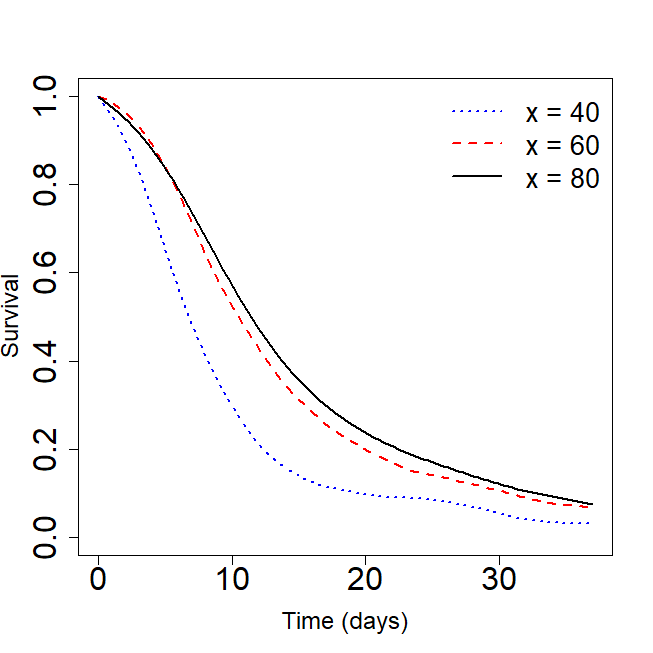}
	
	\caption{Estimation of $S(t|x)$ for time in ward with Beran's estimator (left) and the smoothed Beran's estimator (right) using the bootstrap bandwidths for $x = 40$ (dotted line), $x = 60$ (dashed line) and $x = 80$ (solid line).}
	\label{fig:Bootstrap_S_Time_WARD_bootstrap_Beran_SBeran}
\end{figure}

Figure \ref{fig:Bootstrap_S_Time_WARD_bootstrap_Beran_SBeran} shows the boostrap estimations of the survival function by means of Beran's and the smoothed Beran's estimator. 
The differences between the two estimates lie in the reduction of the roughness of the smoothed Beran estimation. The true survival curve is not expected to exhibit the jumps caused by the classical Beran's estimator. On the contrary, the smoothed Beran estimator returns a smooth curve that, therefore, behaves more similarly to the true survival curve. For this reason, the remainder of this real data analysis is conducted with the smoothed Beran's estimator.

\begin{figure}[H]
	\centering
	\includegraphics[width=0.8\linewidth]{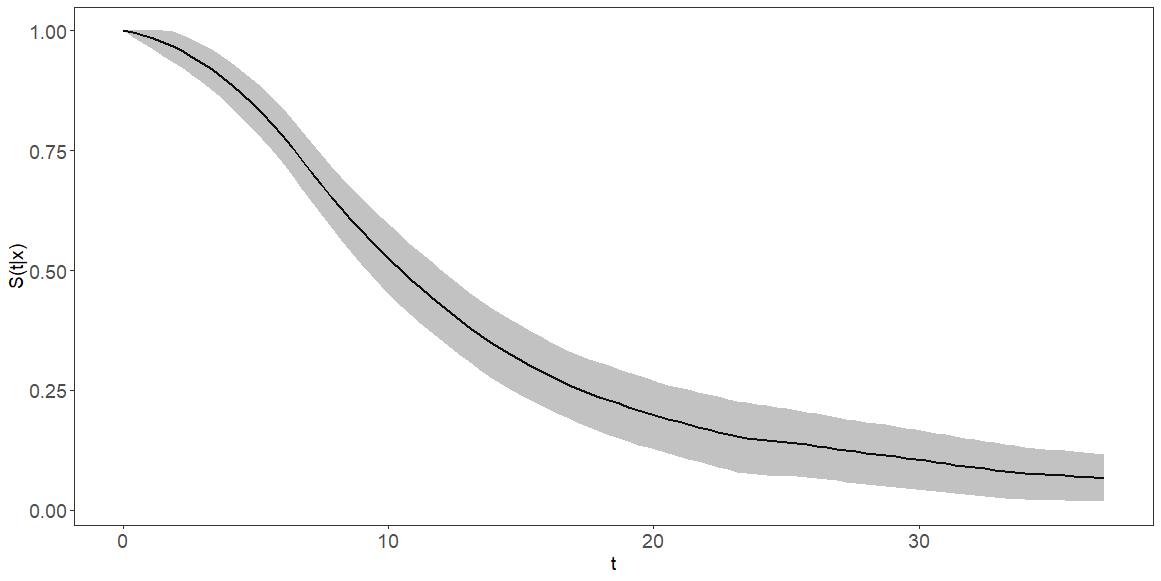}
	\caption{Estimation of $S(t|x)$ with bootstrap bandwidths for time in ward and bootstrap confidence region by means of Method 1 based on the smoothed Beran's estimator for $x = 60$.}
	\label{fig:Bootstrap_S_Time_WARD_bootstrap_SBeran_band_kmethod_60}
\end{figure}

Figure \ref{fig:Bootstrap_S_Time_WARD_bootstrap_Beran_SBeran} states that only $15\%$ of the 40 year old patients spend more than 15 days in ward. Meanwhile, $40\%$ of COVID-19 positive patient of 60 or 80 years old spend more than 15 days in ward and only $20\%$  of these patients spend more than 25 days in ward.
Figure \ref{fig:Bootstrap_S_Time_WARD_bootstrap_SBeran_band_kmethod_60} shows the estimation of the conditional survival function of the time in ward of 60-year-old patients and the bootstrap confidence region with $95\%$ confidence level obtained by Method 1. The average width of the confidence region is $0.1227$.

Length-of-stays were also analysed by classifying individuals into two gender populations. The main conclusions are shown in the following paragraphs.

\begin{figure}[H]
	\centering
	\includegraphics[width=0.45\linewidth]{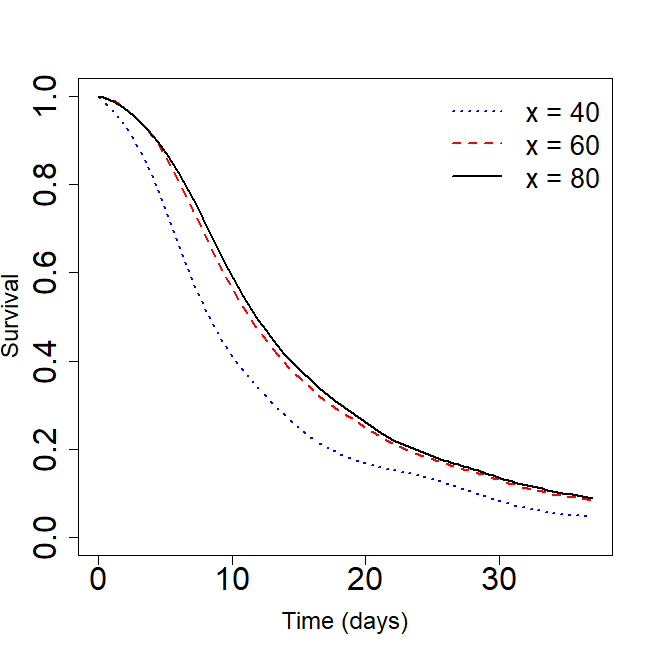}
	\includegraphics[width=0.45\linewidth]{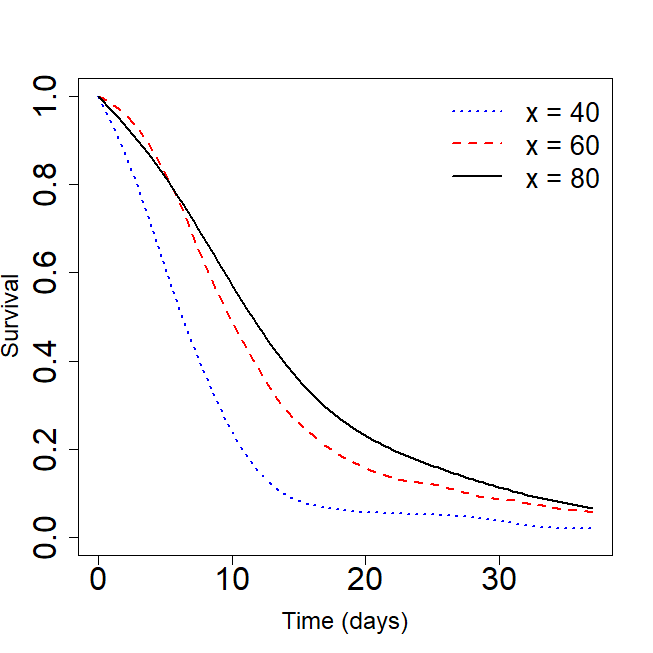}
	
	\caption{Estimation of $S(t|x)$ for time in ward with the smoothed Beran's estimator using the bootstrap bandwidths in the men subpopulation (left) and in the women subpopulation (right) for $x = 40$ (dotted line), $x = 60$ (dashed line) and $x = 80$ (solid line).}
	\label{fig:Bootstrap_S_Time_WARD_bootstrap_SBeran_sex}
\end{figure}

Figure \ref{fig:Bootstrap_S_Time_WARD_bootstrap_SBeran_sex} shows that the differences in length-of-stay between ages when restricted to the men subpopulation are slight. In contrast, the distribution of the time in ward seems to be remarkably different for women of different ages. About $20\%$ of women aged 60-80 spend more than 20 days in ward. Meanwhile, only $10\%$ of 40-year-old women spend more than 20 days in ward. 
Furthermore, this plots show that young women have shorter length-of-stays than young men. On the contrary, differences between male and female populations are insignificant at older ages.

\begin{figure}[H]
	\centering
	\includegraphics[width=0.8\linewidth]{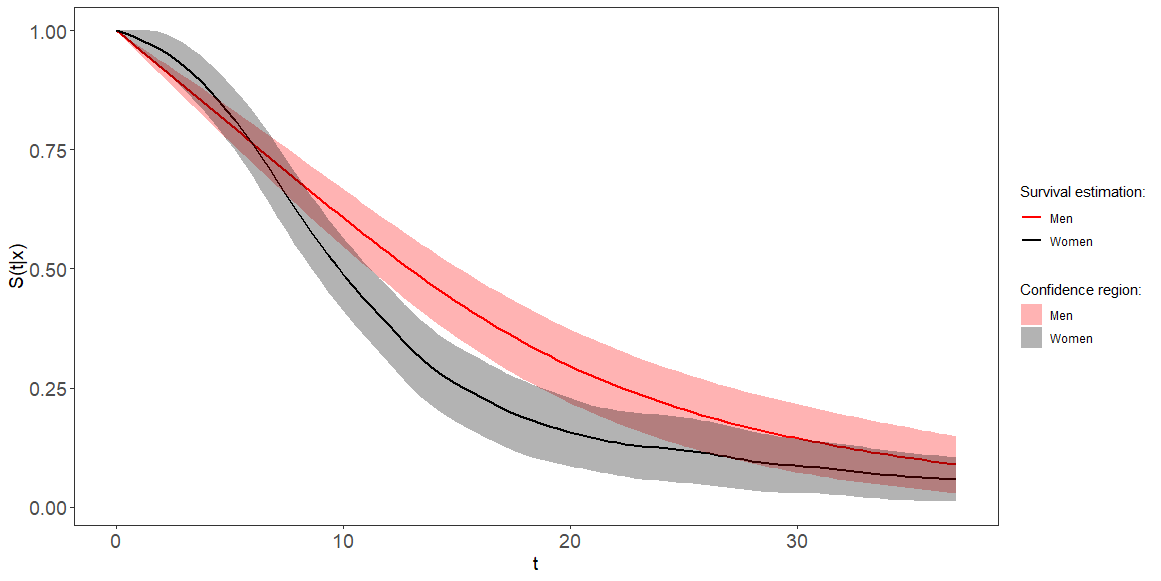}
	\caption{Estimation of $S(t|x)$ with bootstrap bandwidths for time in ward and bootstrap confidence region by means of Method 1 based on the smoothed Beran's estimator for $x = 60$ in the men (red lines) and the women (black lines) subpopulations.}
	\label{fig:Bootstrap_S_Time_WARD_bootstrap_SBeran_band_kmethod_60_sex}
\end{figure}

Figure \ref{fig:Bootstrap_S_Time_WARD_bootstrap_SBeran_band_kmethod_60_sex} shows the estimation of the conditional survival function of the time in ward of 60-year-old men and 60-year-old women and the corresponding bootstrap confidence regions with $95\%$ confidence level. The average width of the confidence region for the men subpopulation is $0.1224$ and for the women subpopulation is $0.1272$. The confidence regions for both men and women subpopulations confirm the observed differences in the length-of-stay in ward between them.

Now, it is considered whether or not patients have COPD. The possible effect of this risk factor on length-of-stays is discussed below.
The age profiles considered here are 60 and 80 years because the proportion of young patients in the sample diagnosed with COPD is low. 

\begin{figure}[H]
	\centering
	\includegraphics[width=0.45\linewidth]{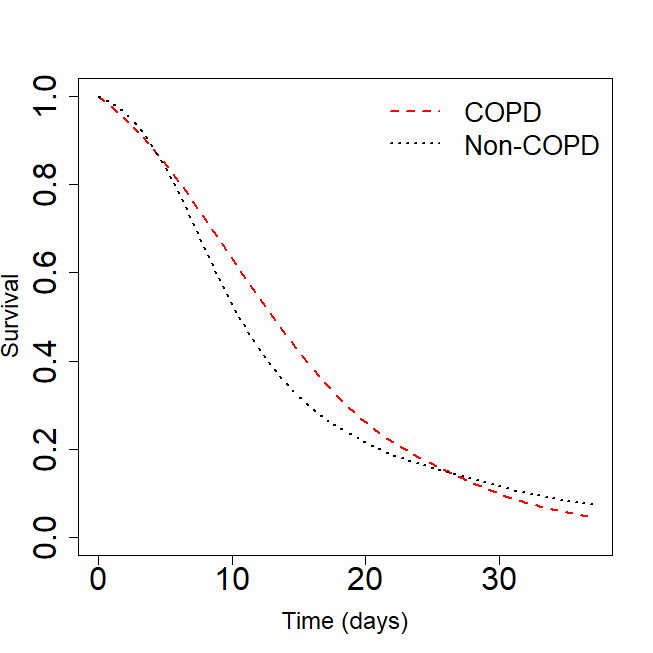}
	\includegraphics[width=0.45\linewidth]{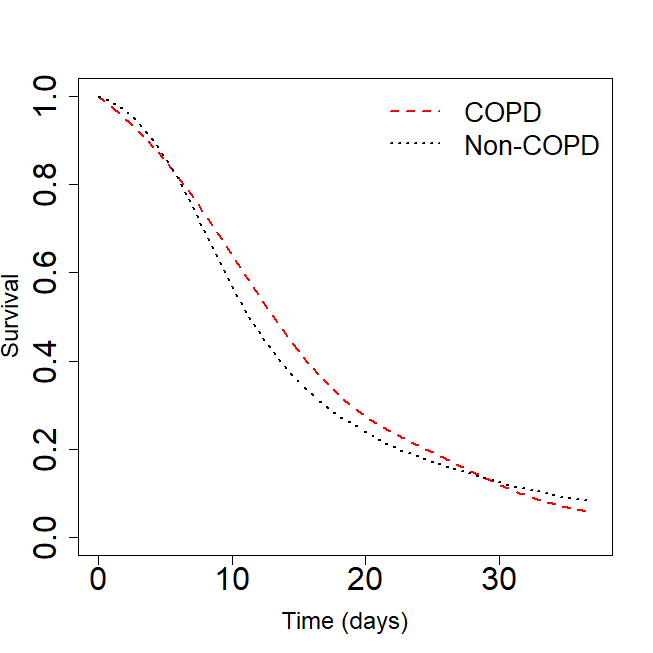}
	
	\caption{Estimation of $S(t|x)$ for time in ward with the smoothed Beran's estimator using the bootstrap bandwidth with $x = 60$ (left) and $x = 80$ (right) in the COPD (dashed lines) and non-COPD (dotted lines) subpopulations.}
	\label{fig:Bootstrap_Time_WARD_bootstrap_SBeran_COPD_age}
\end{figure}

Figure \ref{fig:Bootstrap_Time_WARD_bootstrap_SBeran_COPD_age} shows the survival fucntion of the  length-of-stays of 60-year-old and 80-year-old patients according to whether or not they have COPD. The length-of-stays of COPD patients is slightly higher than in non-COPD patients. The difference is most pronounced in 60-year-old patients.

Another risk factor for COVID-19 disease is obesity, so its possible effect on the length-of-stay in hospital ward is studied.

\begin{figure}[H]
	\centering
	\includegraphics[width=0.45\linewidth]{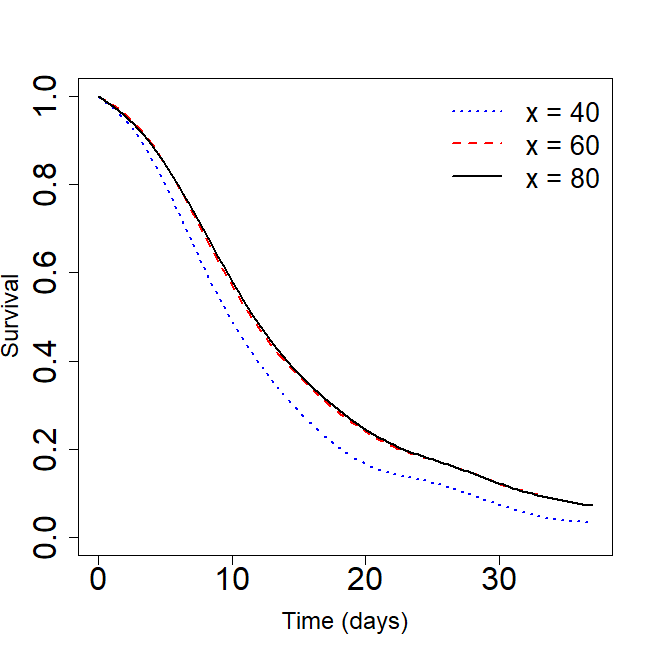}
	\includegraphics[width=0.45\linewidth]{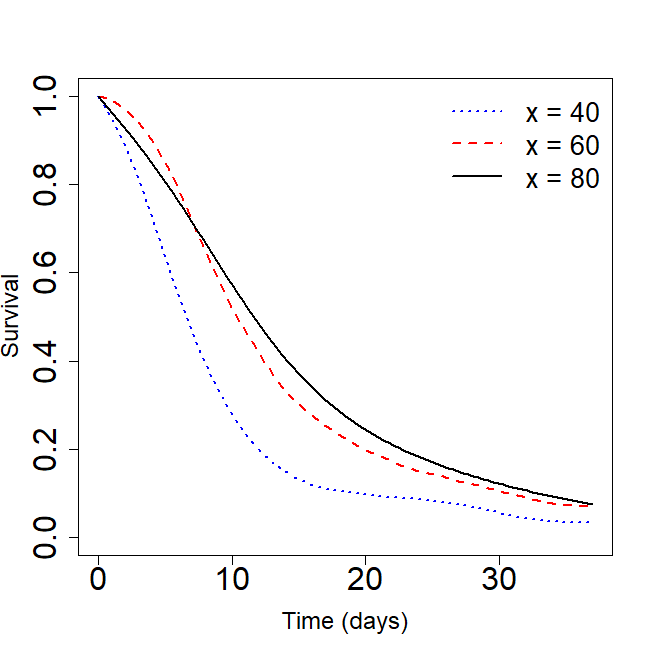}
	
	\caption{Estimation of $S(t|x)$ for time in ward with Beran's estimator using the bootstrap bandwidth in the obesity patients subpopulation (left) and non-obesity patients subpopulation (right) for $x = 40$ (dotted line), $x = 60$ (dashed line) and $x = 80$ (solid line).}
	\label{fig:Bootstrap_S_Time_WARD_bootstrap_SBeran_obesity}
\end{figure}

\begin{figure}[H]
	\centering
	\includegraphics[width=0.45\linewidth]{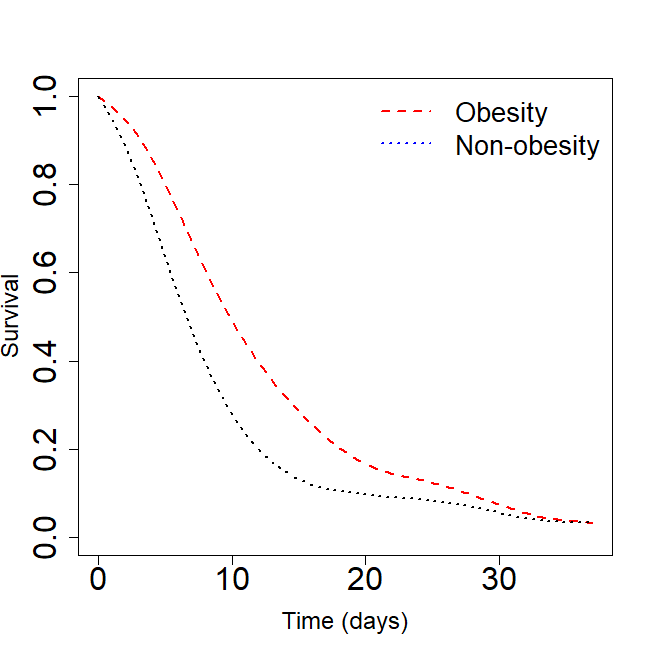}
	\includegraphics[width=0.45\linewidth]{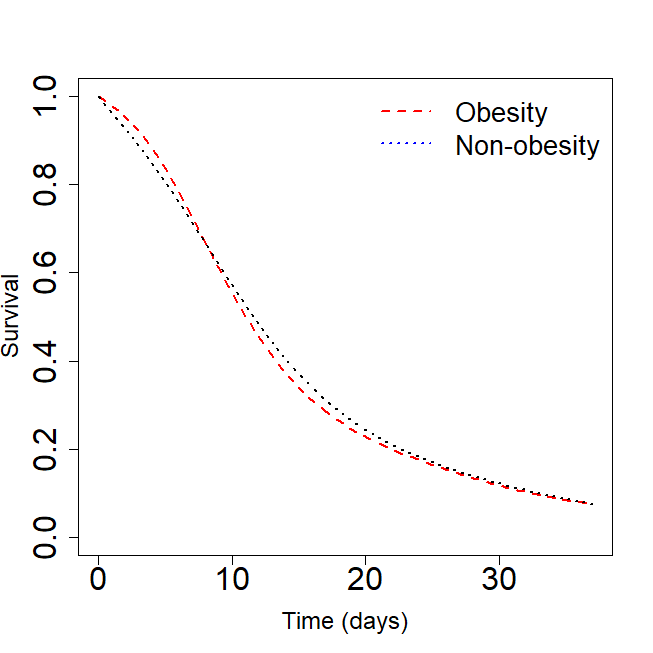}
	
	\caption{Estimation of $S(t|x)$ for time in ward with Beran's estimator using the bootstrap bandwidth with $x = 40$ (left) and $x = 80$ (right) in the obesity (dashed lines) and non-obesity (dotted lines) subpopulations.}
	\label{fig:Bootstrap_S_Time_WARD_bootstrap_SBeran_obesity_age}
\end{figure}

Figure \ref{fig:Bootstrap_S_Time_WARD_bootstrap_SBeran_obesity} shows that the effect of age on length-of-stay is greatly attenuated by obesity. That is, in the case of obesity, the hospitalisation time is similar for all considered ages. Figure \ref{fig:Bootstrap_S_Time_WARD_bootstrap_SBeran_obesity_age} shows that hospitalisation times are somewhat longer in 40-year-old patients with obesity than in 40-year-old patients without obesity. In contrast, at older ages, the effect of this risk factor is not appreciable: hospitalisation times in ward do not differ between patients with and without obesity in their 80s.

\subsection{Length-of-stay in ICU}

The time until leaving the Intensive Care Unit could also be studied. A COVID-19 positive patient leaves the ICU because he/she is discharged (from the ICU) or dies and his/her time until the event of interest is known. In other case, the censoring time is observed. The total number of patients in the ICU who were followed up is 288 and the censoring percentage of this dataset is $14.58\%$. 

The smoothed Beran's estimator is used for estimating the survival function of the time in ICU of the Galician COVID-19 patients with bootstrap bandwidths obtained by the automatic selector proposed in Subsection \ref{Bandwidth_SBeran}. Some interesting confidence regions based on the smoothed Beran's estimator are also obtained. 
The bootstrap estimation is obtained in a grid of time $t_1 < \cdots < t_{n_T}$ with $t_{n_T} = \widehat{Q}(0.95)$ and $n_T = 100$. The pilot bandwidth for the covariate used in the bootstrap algorithm is the one defined in \eqref{eq:Bootstrap_S_piloto_r} with $c = 3/2$. The pilot bandwidth for the time variable was defined in \eqref{eq:Bootstrap_S_piloto_s}.

An analysis of the effect of the age and the factors sex, diagnosis of COPD and obesity, parallel to the one carried out for time in ward, was conducted for time in the ICU. The conclusions of this study for ICU time do not differ from those obtained for ward time. 
Hospitalisation times in the ICU follow a similar pattern to that observed for length-of-stay in the ward. Although this pattern is attenuated in ICU hospitalisations due to the higher severity of all patients considered.
Some relevant remarks are mentioned below.

\begin{figure}[H]
	\centering
	\includegraphics[width=0.45\linewidth]{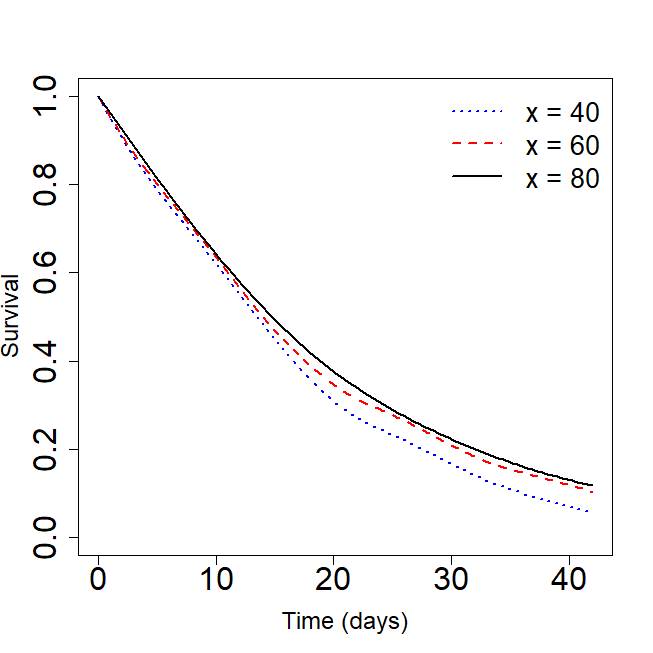}
	\includegraphics[width=0.45\linewidth]{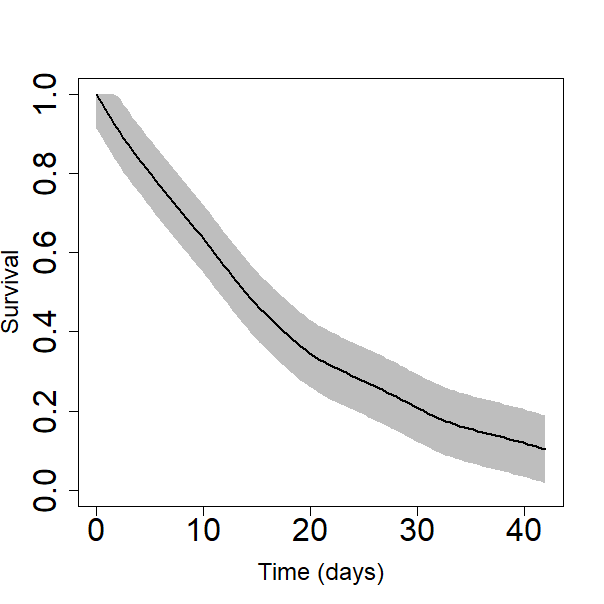}
	\caption{Estimation of $S(t|x)$ for time in ICU with the smoothed Beran's estimator (left) using the bootstrap bandwidths for $x = 40$ (dotted line), $x = 60$ (dashed line) and $x = 80$ (solid line) and the bootstrap confidence region by means of Method 1 for $x = 60$ (right).}
	\label{fig:Bootstrap_S_Time_ICU_bootstrap_SBeran_band}
\end{figure}

Figure \ref{fig:Bootstrap_S_Time_ICU_bootstrap_SBeran_band} shows the survival function of time in the ICU estimated for several ages by means of the smoothed Beran's estimator and the bootstrap confidence region for the conditional survival function of 60-year-old patient with $95\%$ confidence level obtained by Method 1. 
It can be seen, in contrast to the time in ward, that age has little effect on time in the ICU, except in hospitalisations of more than 20 days where slight differences can be seen with time in ICU being shorter in younger age groups.

\begin{figure}[H]
	\centering
	\includegraphics[width=0.8\linewidth]{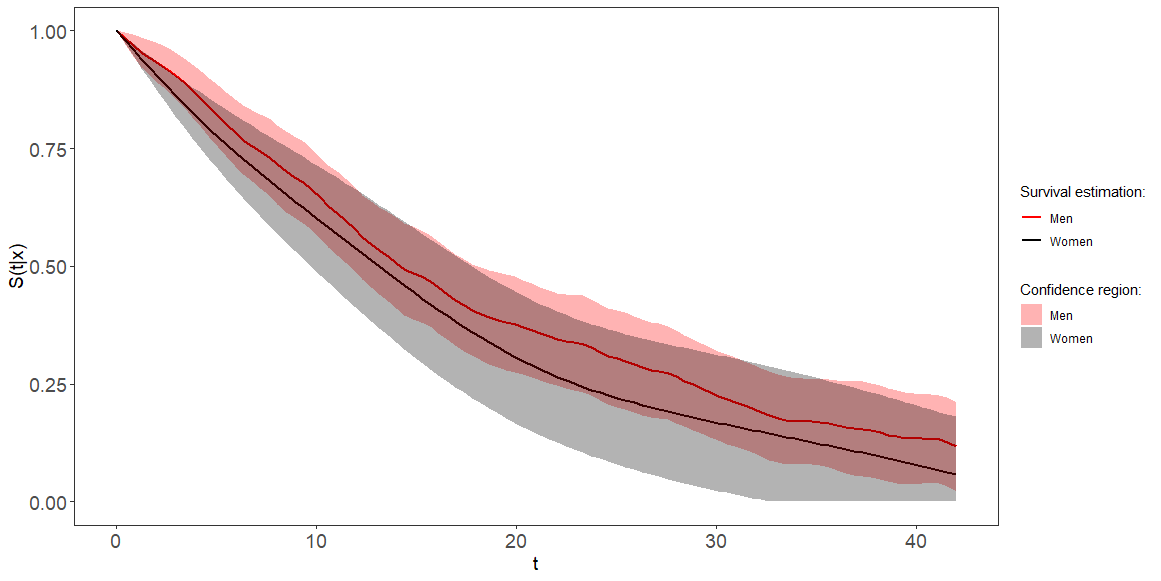}
	\caption{Estimation of $S(t|x)$ with bootstrap bandwidths for time in the ICU and bootstrap confidence region by means of Method 1 based on the smoothed Beran's estimator for $x = 60$ in the men (red lines) and the women (black lines) subpopulations.}
	\label{fig:Bootstrap_S_Time_ICU_bootstrap_SBeran_band_kmethod_60_sex}
\end{figure}

Although age has no global impact on the ICU time, it does have a mild effect when we consider the male and female populations independently. An example of this can be seen in the following plot. Figure \ref{fig:Bootstrap_S_Time_ICU_bootstrap_SBeran_band_kmethod_60_sex} shows the estimation of the conditional survival function of the time in the ICU of 60-year-old men and 60-year-old women and the corresponding bootstrap confidence regions with $95\%$ confidence level obtained by Method 1 based on the smoothed Beran's estimator.

On the other hand, a similar analysis not included here shows that both risk factors, COPD and obesity, have a negative effect on patients' length-of-stay. In both cases this effect is attenuated by age.

\section{Discussion}

The study of the conditional survival function of the variable length-of-stay in hospital ward or ICU for COVID-19 patients admitted in the public hospital system in Galicia since the beginnng of the pandemic until May 11th 2020, carried out in this paper allowed some interesting conclusions to be drawn. 
There are significant  differences in hospitalisation times in ward of COVID-19 patients related to their age. These differences are more accentuated in women.
Although age has no global impact on the length-of-stay, it does have an effect in the female population. Moreover, young women have shorter length-of-stays than young men.
There are differences in length-of-stay in both hospital ward and ICU between patients with or without COPD. The hospitalisation lengths are longer in patients with COPD, although the differences become less significant at older ages.
At younger ages, length-of-stays are significantly longer in patients with obesity. However, the effect of this risk factor is not appreciable at older ages.

The bootstrap selectors of the smoothing parameters provide the possibility to perform this statistical analysis about the length-of-stay of COVID-19 patients, since they allow to obtain reasonable approximations, according to the mean integrated squared error, of the optimal bandwidths involved in Beran's estimator and the doubly smoothed Beran's estimator, which are unknown in practice. Bootstrap-based confidence regions are also informative outputs in this type of analysis.

The proposed methods require a high computational cost. Obtaining the appropriate bootstrap bandwidths to estimate the conditional survival function or computing a confidence region by 500 resamples from a sample of size 1000 requires 3.5 hours, while a sample of size 3000 requires five days. Moreover, these times seem to increase
quadratically as the sample size grows.
For this reason, for considerably large sample sizes, the problem of selecting the smoothing parameters could be addressed by using subsample procedures.

Future work will include the implementation of an R package including the developed software. It will be publicly available on the Comprehensive R Archive Network (CRAN).

\section{Acknowledgements}

This research has been supported by MICINN Grant PID2020-113578RB-100, by the Xunta de Galicia (Grupo de Referencia Competitiva ED431C-2020-14 and Centro Singular de Investigaci\'{o}n de Galicia ED431G 2019/01), all of them through the ERDF.

\bibliographystyle{unsrt}
\bibliography{references}

\begin{thebibliography}{10}

\bibitem{Beran1981}
Rudolf Beran.
\newblock Nonparametric regression with randomly censored survival data.
\newblock {\em Technical report, University of California}, 1981.

\bibitem{KaplanMeier1958}
Edward~L. Kaplan and P.~Meier.
\newblock Nonparametric estimation from incomplete observations.
\newblock {\em Journal of American Statistical Association}, 53(282):457--481,
  1958.

\bibitem{Dabrowska1989}
Dorota~M. Dabrowska.
\newblock Uniform consistency of the kernel conditional {K}aplan-{M}eier
  estimate.
\newblock {\em The Annals of Statistics}, 17(3):1157--1167, 1989.

\bibitem{Manteiga1994}
W.~Gonz\'alez-Manteiga and C.~Cadarso-Su\'arez.
\newblock Asymptotic properties of a generalized {K}aplan-{M}eier estimator
  with applications.
\newblock {\em Nonparametric Statistics}, 4(1):65--78, 1994.

\bibitem{Ingrid1996}
Ingrid Van~Keilegom and N.~Veraverbeke.
\newblock Uniform strong convergence results for the conditional
  {K}aplan-{M}eier estimator and its quantiles.
\newblock {\em Communications in Statistics, Theory Methods}, 25(2):2251--2265,
  1996.

\bibitem{Iglesias1999}
Maria~C. Iglesias-P\'erez and W.~Gonz\'alez-Manteiga.
\newblock Strong representation of a generalized product-limit estimator for
  truncated and censored data with some applications.
\newblock {\em Journal of Nonparametric Statistics}, 10(3):213--244, 1999.

\bibitem{Ingrid1999}
Ingrid Van~Keilegom and M.G. Akritas.
\newblock Transfer of tail information in censored regression models.
\newblock {\em The Annals of Statistics}, 27(5):1745--1784, 1999.

\bibitem{Ingrid2001}
Ingrid Van~Keilegom, M.G. Akritas, and N.~Veraverbeke.
\newblock Estimation of the conditional distribution in regression with
  censored data: a comparative study.
\newblock {\em Computational Statistics and Data Analysis}, 35(4):487--500,
  2001.

\bibitem{Gannoun2005}
Ali Gannoun, Jerome Saracco, A.~Yuan, and G.E. Bonney.
\newblock Nonparametric quantile regression with censored data.
\newblock {\em Scandinavian Journal of Statistics}, 32(4):527--550, 2005.

\bibitem{Gannoun2007}
Ali Gannoun, Jerome Saracco, and Keming Yu.
\newblock Comparison of kernel estimator of conditional distribution function
  and quantile regression under censoring.
\newblock {\em Statistical Modelling}, 7(4):329--344, 2007.

\bibitem{Cai2003}
Zongwu Cai.
\newblock Weighted local linear approach to censored nonparametric regression.
\newblock In Michael~G. Akritas and Dimitris~N. Politis, editors, {\em Recent
  Advances and Trends in Nonparametric Statistics}, pages 217--231. Elsevier,
  Amsterdam, 2003.

\bibitem{Pelaez2022_JNPS}
R.~Pel\'aez, R~Cao, and J.M. Vilar.
\newblock Nonparametric estimation of the conditional survival function with
  double smoothing.
\newblock {\em Journal of Nonparametric Statistics}, DOI:
  10.1080/10485252.2022.2102631, 2022.

\bibitem{Efron1981}
B.~Efron.
\newblock Censored data and the bootstrap.
\newblock {\em Journal of American Statistical Association}, 76(374):312--319,
  1981.

\bibitem{Akritas1986}
M.~Akritas.
\newblock Bootstrapping the {K}aplan-{M}eier estimator.
\newblock {\em Journal of American Statistical Association},
  81(396):1032--1039, 1986.

\bibitem{Lo_Singh1986}
S.~H. Lo and K.~Singh.
\newblock The product-limit estimator and the bootstrap: Some asymptotic
  representations.
\newblock {\em Probability Theory and Related Fields}, 71(3):455--465, 1986.

\bibitem{Ingrid1997}
Ingrid Van~Keilegom and Noel Veraverbeke.
\newblock Estimation and bootstrap with censored data in fixed design
  nonparametric regression.
\newblock {\em Annals of the Institute of Statistical Mathematics},
  49(3):467--491, 1997.

\bibitem{Li_Datta2001}
Gand Li and Somnath Datta.
\newblock A bootstrap approach to nonparametric regression for right censored
  data.
\newblock {\em The Institute of Statistical Mathematics}, 53(4):708--729, 2001.

\bibitem{Geerdens2017}
C.~Geerdens, E.~F. Acar, and P.~Janssen.
\newblock Conditional copula models for right-censored clustered event time
  data.
\newblock {\em Biostatistics}, 19(2):247--262, 2017.

\bibitem{Pelaez2022_Mathematics}
R.~Pel\'aez, R~Cao, and J.M. Vilar.
\newblock Bootstrap bandwidth selection and confidence regions for double
  smoothed default probability estimation.
\newblock {\em Mathematics}, 10(9):1523, 2022.

\bibitem{Pelaez2021_TEST}
R.~Pel\'aez, R~Cao, and J.M. Vilar.
\newblock Probability of default estimation in credit risk using a
  nonparametric approach.
\newblock {\em TEST}, 30(2):383--405, 2021.

\bibitem{Pelaez2021_SORT}
R.~Pel\'aez, R~Cao, and J.M. Vilar.
\newblock Nonparametric estimation of probability of default with double
  smoothing.
\newblock {\em SORT}, 45(2):93--120, 2021.

\bibitem{Azzalini1981}
A.~Azzalini.
\newblock A note on the estimation of a distribution function and quantiles by
  a kernel method.
\newblock {\em Biometrika}, 68(1):326--328, 1981.

\bibitem{Jones1990}
M.~C. Jones.
\newblock The performance of kernel density functions in kernel distribution
  function estimation.
\newblock {\em Statistics and Probability Letters}, 9(2):129--132, 1990.

\bibitem{Foldes1981}
A.~F\"{o}ldes, L.~Rejt\o, and B.~B. Winter.
\newblock Strong consistency properties of nonparametric estimators for
  randomly censored data, ii: estimation of density and failure rate.
\newblock {\em Periodica Mathematica Hungarica}, 12(1):15--29, 1981.

\bibitem{Leconte2002}
Eve Leconte, Sandrine Poiraud-Casanova, and Christine Thomas-Agnan.
\newblock Smooth conditional distribution function and quantiles under random
  censorship.
\newblock {\em Lifetime Data Analysis}, 8(3):229--246, 2002.

\bibitem{Billingsley1968}
P.~Billingsley.
\newblock Convergence of probability measure. {W}iley {S}eries in {P}robability
  and {M}athematical {S}tatistics: Tracts on probability and statistics.
\newblock {\em John Wiley and Sons, New York}, 9, 1968.

\bibitem{optim1995}
Richard~H. Byrd, Peihuang Lu, Jorge Nocedal, and Ciyou Zhu.
\newblock A limited memory algorithm for bound constrained optimization.
\newblock {\em SIAM Journal on Scientific Computing}, 16(5):1190--1208, 1995.

\bibitem{optim1997}
Ciyou Zhu, Richard~H. Byrd, Peihuang Lu, and Jorge Nocedal.
\newblock Algorithm 778: {L-BFGS-B}: Fortran subroutines for large-scale
  bound-constrained optimization.
\newblock {\em ACM Transactions on Mathematical Software}, 23(4):550--560,
  1997.

\bibitem{Cao2010}
Ricardo Cao, M.~Francisco-Fern\'{a}ndez, and E.J. Quinto.
\newblock A random effect multiplicative heteroscedastic model for bacterial
  growth.
\newblock {\em BMC Bioinformatics}, 11(77), 2010.

\bibitem{Zhun2017}
T.~Zhun and D.N. Politis.
\newblock Kernel estimates of nonparametric functional autoregression models
  and their bootstrap approximation.
\newblock {\em Electronic Journal of Statistics}, 11(2):2876--2906, 2017.

\bibitem{Winkler1972}
Robert~L. Winkler.
\newblock A decision-theoretic approach to interval estimation.
\newblock {\em Journal of the American Statistical Association},
  67(337):187--191, 1972.

\end{thebibliography}

\end{document}